\documentclass{nature1}

\usepackage{graphicx}
\usepackage{epsfig}
\usepackage{amssymb}
\usepackage{amsmath}
\usepackage{mathrsfs}
\usepackage{color}

\newcommand{\aap}{{\it Astron. Astrophys.}}
\newcommand{\araa}{{\it Ann. Rev. Astron. Astrophys.}}
\newcommand{\apj}{{\it Astrophys. J.}}
\newcommand{\aj}{{\it Astron. J.}}
\newcommand{\apjl}{{\it Astrophys. J. Letters}}
\newcommand{\apjs}{{\it Astrophys. J. Supp.}}
\newcommand{\nat}{{\it Nature}}
\newcommand{\newasr}{{\it New Astron. Rev.}}

\newcommand{\mnras}{{\it Mon. Not. R. Astron. Soc.}}
\newcommand{\apss}{{\it Astrophys. Space Sci.}}

\def\bhm{M_{\bullet}}

\def\etal{{\it et al.}}

\def\ergs{\rm erg~s^{-1}}

\def\kms{\rm km~s^{-1}}

\def\Loiii{L_{\rm [OIII]}}

\def\mathdotM{\dot{\mathscr{M}}}

\def\rblr{R_{_{\rm BLR}}}

\def\Rg{R_{\rm g}}
\def\Rsg{R_{\rm sg}}

\def\sunm{M_{\odot}}

\def\vlos{v^{_{^{_{\|}}}}}

\def\gammaA{\gamma_{_{\rm A}}}
\def\gammaB{\gamma_{_{\rm B}}}
\def\gammaC{\gamma_{_{\rm C}}}

\def\calA{{\cal A}_{\rm H\beta}}
\def\calS{{\cal S}_{\rm H\beta}}
\def\calZ{{\cal Z}_{\rm H\beta}}

\def\civ{C {\sc iv}}
\def\feii{Fe {\sc ii}}

\def\oiii{[O~{\sc iii}]}

\def\cblue{\color{black}}

\def\vXC{v_{_{\XC}}}
\def\vYC{v_{_{\YC}}}
\def\vZC{v_{_{\ZC}}}

\def\XC{X_{_{\rm C}}}
\def\YC{Y_{_{\rm C}}}
\def\ZC{Z_{_{\rm C}}}

\def\varphib{\varphi_{\rm b}}

\def\alphaC{\alpha_{_{\rm C}}}

\def\thetaC{\theta_{_{\rm C}}}

\def\varphiC{\varphi_{_{\rm C}}}

\def\Dtid{D_{\rm tid}}
\def\Dsub{D_{\rm sub}}
\def\NH{N_{\rm H}}

\def\alphaA{\alpha_{_{\rm A}}}
\def\betaA{\beta_{_{\rm A}}}

\begin{document}

\title{Tidally disrupted dusty clumps as the origin of broad emission lines in active galactic nuclei}
\author{
Jian-Min Wang$^{1,2,3}$, Pu Du$^{1}$, Michael S. Brotherton$^{4}$, Chen Hu$^{1}$,
Yu-Yang Songsheng$^1$, \\ Yan-Rong Li$^{1}$, Yong Shi$^{5}$ and Zhi-Xiang Zhang$^{1}$
}
\maketitle

\begin{affiliations}
\item{Key Laboratory for Particle Astrophysics, Institute of High Energy Physics,
Chinese Academy of Sciences, 19B Yuquan Road, Beijing 100049, China}

\item{School of Astronomy and Space Science, School of Physical Sciences, University of Chinese Academy of Sciences, 
19A Yuquan Road, Beijing 100049, China}  

\item{National Astronomical Observatories of China, Chinese Academy of Sciences,
 20A Datun Road, Beijing 100020, China}
 
\item{Department of Physics and Astronomy, University of Wyoming, Laramie, WY 82071, USA}

\item{School of Astronomy and Space Science, Nanjing University,
Nanjing 210093, China}
\end{affiliations}

\begin{abstract}
Type 1 active galactic nuclei display  broad emission lines, regarded as arising from 
photoionized gas moving in the gravitational potential of a supermassive black 
hole\cite{Ho2008,Gaskell2009}. The origin of this broad-line region gas is unresolved 
so far\cite{Ho2008,Gaskell2009,Korista1999}, however. Another component is the dusty 
torus\cite{Antonucci1993} beyond the broad-line region, likely an assembly of discrete 
clumps\cite{Jaffe2004,Elitzur2006,Nenkova2008} that can hide the region from some 
viewing angles and make them observationally appear as Type 2 objects. Here we report 
that these clumps moving within the dust sublimation radius, like the molecular cloud 
G2 discovered in the Galactic center\cite{Gillessen2012}, will be tidally disrupted 
by the hole, resulting in some gas becoming bound at smaller radii while other gas 
is ejected and returns to the torus. The clumps fulfill necessary conditions to be 
photoionized\cite{Osterbrock2006}. Specific dynamical components of tidally disrupted 
clumps include spiral-in gas as inflow, circularized gas, and ejecta as outflow. We 
calculate various profiles of emission lines from these clouds, and find they generally 
agree with H$\beta$ profiles of Palomar-Green quasars\cite{Boroson1992}. We find that 
asymmetry, shape and shift of the profiles strongly depend on \oiii\, luminosity, 
which we interpret as a proxy of dusty torus angles. Tidally disrupted clumps from 
the torus may represent the source of the broad-line region gas.
\end{abstract}

The mass of individual clumps can be estimated by the tidal disruption condition at the dust 
sublimation radius given by the inner edge of torus,
$\Dsub\approx 0.4\,L_{45}^{1/2}T_{1500}^{-2.6}\,{\rm pc}$, where $L_{45}=L_{\rm UV}/10^{45}\,\ergs$
and $T_{1500}=T_{\rm sub}/1500$K is the dust sublimation 
temperature\cite{Barvainis1987,Elitzur2006,Laor1993}. The tidal disruption happens at or inside 
the Roche limit of $\Dtid=\left(\bhm/M_{\rm C}\right)^{1/3}R_{\rm C}$, where $\bhm$ is the black 
hole mass, $M_{\rm C}$ is the clump mass, and $R_{\rm C}=\left(M_{\rm C}/\pi\NH m_{\rm p}\right)^{1/2}$ 
is the size, where $\NH$ is the column density and $m_{\rm p}$ is the proton mass. 
Dust-free clouds are disrupted when $D_{\rm tid}\lesssim D_{\rm sub}$, and therefore the largest
dust-free clouds are given by
$M_{\rm C}/M_{\oplus}\approx 2.4 \,\epsilon^3M_8T_{1500}^{-15.6}N_{24}^3$,
their size is 
$R_{\rm C}=5.1\times 10^{13}\,\epsilon^{3/2}M_8^{1/2}N_{24}T_{1500}^{-7.8}\,$cm
and density 
$n_{\rm C}=1.5\times 10^{10}\,\epsilon^{-3/2}M_8^{-1/2}T_{1500}^{7.8}\,{\rm cm^{-3}}$,
where $M_{\oplus}$ is the earth mass, $\epsilon=L_{45}/M_8$, $M_8=\bhm/10^8\sunm$ and 
$N_{24}$ = $\NH$/10$^{24}$ cm$^{-2}$.  For active galactic nuclei (AGNs) with 
$\bhm/\sunm=10^6\sim10^9$ and $\epsilon=1$, we have $M_{\rm C}/M_{\oplus}\approx 0.02\sim24.0$. 
This estimate shows that the typical properties of the captured 
clumps generally fulfill the photoionization condition for broad emission lines 
of AGNs and quasars\cite{Osterbrock2006}. Moreover, we realize that ($M_{\rm C},R_{\rm C},n_{\rm C}$)
are very sensitive to $T_{\rm sub}$, implying that properties of broad-line regions (BLR) generally 
depend on processes of dust production in galactic nuclei. Note that $\epsilon=1$ corresponds to 
Eddington ratios of $L_{\rm Bol}/L_{\rm Edd}\approx 0.35$, where $L_{\rm Bol}\approx 5L_{\rm UV}$ 
is used\cite{Elvis1994}. In the Galactic center, G2 with a mass of 
$3M_{\oplus}$ captured by the central black hole\cite{Gillessen2012} is well in this range of 
$M_{\rm C}$, lending support for the idea that such captures could be common in other galactic 
centers. 

{\color{black}Collisions among clumps drive them to fall within $\Dsub$ and 
determine tidal capture rates. For a simple model\cite{Krolik1988}, the rates are about
$\dot{M}\sim 1.0{\sunm\rm\ yr^{-1}}$ 
for typical torus parameters,
which is a typical AGN accretion rate, implying a promising fueling mechanism.}
The stationary numbers of the captured clumps can be evaluated by continuity from torus to
the BLR. The number density of captured clouds ($N_{\rm C}$) can be simply estimated from 
the fuelling rates, which is written as
$\dot{M}\approx 2\Delta\Omega D_{\rm sub}^2N_{\rm C}M_{\rm C}f_{v}V_{\rm ff}$,
where $\Delta \Omega=2\pi \sin\Theta_{\rm torus}$ 
is the half solid angle subtended by the torus
(see Figure 1), $V_{\rm ff}=\left(G\bhm/D_{\rm sub}\right)^{1/2}$ is the free-fall velocity
and $f_{v}$ is a fraction of the free-fall velocity.  
The total number of clouds is about $N_{\rm tot}=2\Delta\Omega D_{\rm sub}^3N_{\rm C}/3$, 
leading to \\
$N_{\rm tot}\approx 6\times 10^7\, \varepsilon_0f_{0.3}^{-1}M_8^{1/4}N_{24}^{-3}T_{1500}^{11.7}$,
where $\varepsilon_0=\epsilon^{-9/4}\dot{M}_1M_8^{-1}$, $f_{0.3}=f_{v}/0.3$ and
$\dot{M}_1=\dot{M}/1\sunm{\rm yr^{-1}}$ supplied typically by the torus\cite{Krolik1988}. For 
typical values of parameters, this number is well in agreement with $3\times 10^7$ clouds in 
NGC 4151, the lower limit based on Keck high-resolution spectroscopic observations\cite{Arav1998}. 
If $T_{\rm sub}\sim 2000$K, $N_{\rm tot}\sim 1.7\times 10^9$.  In principle, tidal captures by
the central black hole can generally supply enough clouds to the BLR.

Capture rates are 
$\dot{{\cal R}}\sim \dot{M}/M_{\rm C}\approx 10^5\,\dot{M}_1M_{\rm C,2.4}^{-1}{\rm yr^{-1}}$,
where $M_{\rm C,2.4}=M_{\rm C}/2.4M_{\oplus}$. Such a high rate makes capture events essentially 
continuous to form stationary spiral inflows to the black hole. Captured clumps at the mid-plane 
will naturally form and fuel accretion discs. However, vertical self-gravity instabilities drive 
the outer regions of the discs, beyond\cite{Laor1989}
$\Rsg/\Rg\approx 603\,\left(\alpha_{0.1}/M_8\right)^{2/9}\mathdotM^{4/9}$, 
to fragment into dense clouds or stars, where $\Rg=G\bhm/c^2$ 
is the gravitational radius, $\alpha_{0.1}=\alpha_{_{\rm SS}}/0.1$ is the viscosity parameter, 
and $\mathdotM=\dot{M}/\dot{M}_{\rm Edd}$ is the dimensionless rate normalized by 
$\dot{M}_{\rm Edd}=0.22\,M_8\,\sunm{\rm yr^{-1}}$. Such a self-gravitating disc
is composed of dense clouds and diffuse dusty gas\cite{Begelman1989,Collin1999}.
The disc conditions allow the BLR clouds to pass through, but the optical depth of the 
gas, $\tau\approx 4.0(\lambda/0.7\mu{\rm m})^{-1.5}N_{22}$ for Milky Way dust\cite{Draine2003}, 
is sufficient to cause line emission from the BLR clouds through the disc to experience high 
extinction, where $N_{22}=N_{\rm H}^{\rm d}/10^{22}{\rm cm^{-2}}$ is hydrogen column density 
of the disc. The captured clouds from the lower half-plane torus are obscured by the
disc, resulting in asymmetries of line profiles. 
There may be a fraction of clumps colliding with the dense clouds of the self-gravitating disc
beyond $\Rsg$, but most of them can penetrate the disc between $R_{\rm BLR}\in [\Rsg,D_{\rm sub}]$, 
where $R_{\rm BLR}/\Rg\approx 6.2\times 10^3\,L_{44}^{1/2}M_8^{-1}$ is the emissivity-averaged 
BLR radius\cite{Bentz2013} and $L_{44}$ is the 5100\AA\, luminosity in units of $10^{44}\ergs$.

Unlike the tidal disruption of stars\cite{Rees1988}, 
the captured clumps are undergoing a more complex process, governed additionally by other 
significant factors such as ram pressure, ablation due to friction, and evaporation from thermal 
conduction with a hot medium and drag forces of the surroundings\cite{Burkert2012}. The fates of 
the clumps are difficult to assess due to uncertainty as to their internal states but they are either
distorted or partially disrupted. See a brief discussion on this issue given in the Methods section. 
However, there is evidence that cloud G2 could be connected with G1\cite{Pfuhl2015,McCourt2015}, 
implying that the clumps may split as a result of the tidal force. If the clumps are supported 
by an internal magnetic field at a level of milli Gauss\cite{Krolik1988,Pfuhl2015,McCourt2015}, 
their fates  will be similar to the case of stars: 
the tidal torque spins them up and imparts the orbital energy to the captured clumps until they
split into two parts of unequal energy, one bound and the other unbound\cite{Rees1988}. A fraction 
of the orbital energy is thus channelled into the ejection of gas. The ejection fraction ($f_{\rm C}$) 
of the captured clumps depends on their internal states, and we estimate this dependence in the 
Methods section. Figure 1{\it a} shows a cartoon of the present scenario.

Generally, the bound clouds experience two phases: 1) spiral-in with an elliptical orbit 
and 2) circularization driven by the factors mentioned previously\cite{Burkert2012,McCourt2015}. 
We refer to them as type A and B clouds, respectively.
Here circularized clouds mean their rotation dominates over their
infalling velocity. Type B clouds eventually merge into the accretion disc, otherwise, there is
a pile-up of material from the tidal captures. The unbound material, type C clouds, are ejected. 
We can broadly characterize the type A, B and C clouds as infalling, rotating, and outflowing, 
respectively.

We now present the profiles of these tidally disrupted clumps under the frame as shown by 
Figure 1{\it b}. All quantities are explained in Supplementary Table 1 in the Methods section. 
For type A clouds, ($\zeta_0,\xi_{\rm A},\gammaA,i,\Theta_{\rm torus}$) are the main determinants 
of their contribution to the emission-line profile. $\zeta_0$ determines trajectories of type 
A clouds and hence their velocity structure. Fixing 
$(\xi_{\rm A},\gammaA,i,\Theta_{\rm torus})=(0.9,0,60^{\circ},70^{\circ})$, 
we show the dependence of the 
resulting profiles on $\zeta_0$ in Figure 2{\it a}. Clouds with small $\zeta_0$ will fall
rapidly inward, for example, ones with $\zeta_0=0.1$, infall dominates their dynamics. 
Profiles are dominated by the infalling component and generally show redshifted peaks with 
$\zeta_0$ as shown by Figure 2{\it a}. Decreasing $\zeta_0$, profiles change in two ways: 
1) the red peak shifts increasingly redward and 2) asymmetry is decreased. Large$-\zeta_0$ 
clouds spiral in slowly, over many orbits, showing quasi-symmetric profiles similar to type 
B clouds (i.e., more components from rotation-dominated clouds). These calculations show 
that tidally disrupted clumps observationally appear as asymmetric velocity profiles with 
centroid shifts.

Figure 2{\it b} shows the profile dependence on $\gammaA$ for 
$(\zeta_0,\xi_{\rm A},i,\Theta_{\rm torus},)=(1,0.9,60^{\circ},70^{\circ})$. 
The number of high-speed clouds increases if $\gammaA$ increases, and profiles 
shift toward the red. Figure 2{\it c} shows the dependence on $\Theta_{\rm torus}$. Small 
$\Theta_{\rm torus}$ makes the capture plane more likely to have large $i$ as viewed by 
observers, decreases the projected velocities of the infall and hence the red wings. Double 
peaks appear for cases with a geometrically thin torus  (i.e., small $\Theta_{\rm torus}$) 
and low-$i$ observers. This has important implications, 
namely that AGNs with large torus angles will have an excess of red emission relative to 
the blue. The dependence of profiles on orientation is shown by Figure 2{\it d}. It is 
clear that the more face-on system leads to narrower profiles as 
discovered, but asymmetries remain, with excess emission on the redshifted side of the 
profile. Profiles also depend on $\xi_{\rm A}$, but only linearly.

For type B clouds, the profiles, which are relatively simple compared to those from type A 
clouds, are mainly determined by the inner radius ($R_{\rm in}$) and the spatial distribution 
of clouds along the radius $N_{\rm B}\propto v_{\rm B}^{\gammaB}$ as well as $i$, 
$N_{\rm B}$ is the number of clouds per unit radius, and $\gammaB$ is an index. 
We take $r_0=R_{\rm BLR}/R_{\rm T}^0=0.62$ for the circularized radius.
Figure 2{\it e} and 2{\it f} show the spectral dependences 
on $\gammaB$ and $i$, respectively. Given an $i$, the profiles broaden with 
increasing $\gammaB$ since more clouds are located in the inner regions. The higher the 
inclination $i$, the narrower the profiles.  Type B clouds may contribute the majority 
of the total BLR emission, but they could be mixed with the type A profiles with 
high-$\zeta_0$. 

Generally, type C clouds simply show blue-shifted emission lines as in Figures 2{\it g} 
and 2{\it h}. Except for inclination, the ejection velocity and $\gammaC$ are the main 
parameters governing the profiles emitted by these clouds. The larger $\gammaC$, the less 
blue-shifted the profile  because there are fewer high speed clouds than the low speed. 
The blue peak shifts with inclination simply due to the projection of the net velocity 
of the clouds. Obviously, the clouds contribute to the blue excess of the total profiles 
shown by Figure 2{\it g} and 2{\it h} for different parameters.

Emission from type A, B and C clouds with different ratios and shifts generate a diverse 
range of BLR emission-line profiles. We took efforts to 
calculate profiles for a larger ranges of model parameters than that shown in Figure 2 
in order to find the nature of the dependence of the profiles on each parameter. From these
exercises, we in the end chose 10 as primary parameters and others as the auxiliary (the 
latter of which we held fixed in the fittings for all objects). We apply the present model 
to PG quasars to determine if it can account for the nature of the observed H$\beta$ 
profiles (see the fitting scheme explained in the Methods). 

Figure 3 shows fitting results of H$\beta$ profiles in four example quasars (parameters
are provided by Supplementary Table 2); a complete sample of PG quasar H$\beta$ profile 
fittings are provided in Methods, including the distributions of fitting parameters. 
Supplementary
Table 3 lists classifications of the resultant fittings according to the relative fluxes 
of type A, B, and C clouds to the total.  We find that the average fraction of type C clouds 
is $\langle f_{\rm C}\rangle\approx 0.1$ and $f_{\rm C}\ll(f_{\rm A},f_{\rm B})$ in
most objects from their distributions in the Method.  This nicely agrees with the estimate 
of Equation (\ref{fc}). The fittings indicate that H$\beta$ profiles 
of PG quasars can be generally fitted very well, showing robustness of the present model.
Moreover, the fittings decompose entire H$\beta$ profiles for physical analyses.

Defining $\calA$, $\calS$ and $\calZ$ as asymmetries, shapes, and shifts of profiles 
in Methods, we explore their correlations with the observed \oiii$\lambda5007$ luminosity 
$(\Loiii)$ as a proxy of dusty torus angles\cite{Simpson2005,Maiolino2007,Reyes2008}. 
Correlations are shown in Figure 4. Comparing with correlations found in PG quasars 
(see Table 3 in Ref.\cite{Boroson1992}), we find that the present correlations are among 
the strongest. Therefore, properties of the physically decomposed components ($\calA$, 
$\calS$ and $\calZ$) strongly depend on $\Loiii$, and likely torus angles, lending support 
for the origin of the BLR from the torus.

The possibility of discrete clouds as the BLR model, supported by 
observations\cite{Risaliti2007}, has been suggested for many years\cite{Krolik1981}, 
but their origin and supply remain open questions\cite{Mathews1986} so far. 
In this paper, we demonstrate one possibility of the origin of the BLR. The simplified 
model yields major features of observational characteristics. In the future, it  
is worth using numerical simulations to investigate the entire dynamical evolution 
of the clumps which would provide more detailed comparisons with observed profiles. 
In such a model, the BLR is one phase in the path of fueling the black hole, physically 
connecting the torus, BLR, and accretion discs to each other. A self-consistent
model should consider the dynamics and thermodynamics of clouds not only for profiles, 
but also for low/high ionisation regions emitting different lines. Reverberation of 
H$\beta$ line and near-infrared emissions with respect to the varying continuum in AGNs, 
in particular ones with asymmetric H$\beta$ profiles, will provide tests of the connection 
between the BLR and torus components. Given the importance of 
broad-emission lines in measuring black hole masses, this is a compelling goal.

\clearpage
 
\begin{addendum}

\item 
The authors are grateful to three anonymous referees for helpful reports improving the manuscript.
Todd Boroson is thanked for sending the data of PG quasar spectra. This research 
is supported by National Key Program for Science and Technology Research and Development (grant 
2016YFA0400701) and  grants NSFC-11173023, -11133006, -11373024, 
-11233003 and -11473002, and by Key Research Program of Frontier Sciences, CAS, Grant 
QYZDJ-SSW-SLH007.

\item[Author Contributions] 
JMW conceived the project through presenting the idea and building up the 
current model. JMW and DP jointly made calculations. JMW and MSB jointly wrote the manuscript. 
YYS and JMW fitted H$\beta$ profiles of PG quasars, CH, ZXZ and YS measured PG quasar spectra.
JMW and YRL shared many discussions on clump physics. All the authors discussed the contents 
of the paper.

\item[Correspondence] 
Correspondence and requests for materials
should be addressed to Jian-Min Wang (email: wangjm@ihep.ac.cn).

\item[Competing Interests] The authors declare that they have no
competing financial interests.

\end{addendum}

\newpage

\begin{figure}
\centering
\includegraphics[angle=0,origin=c,trim=0pt 90pt 0pt 80pt, width=0.58\textwidth]{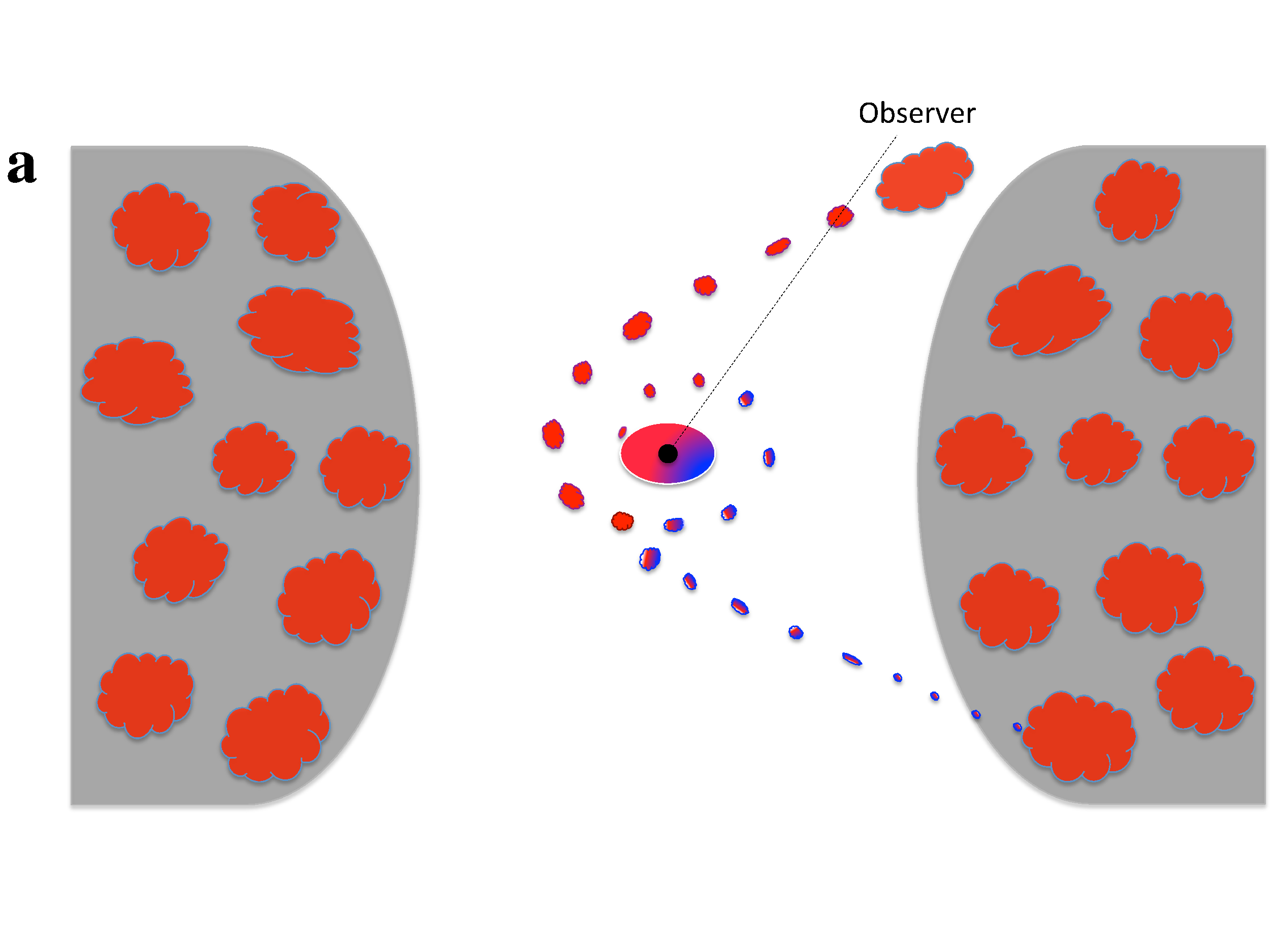}
\includegraphics[angle=0,origin=c,width=0.41\textwidth]{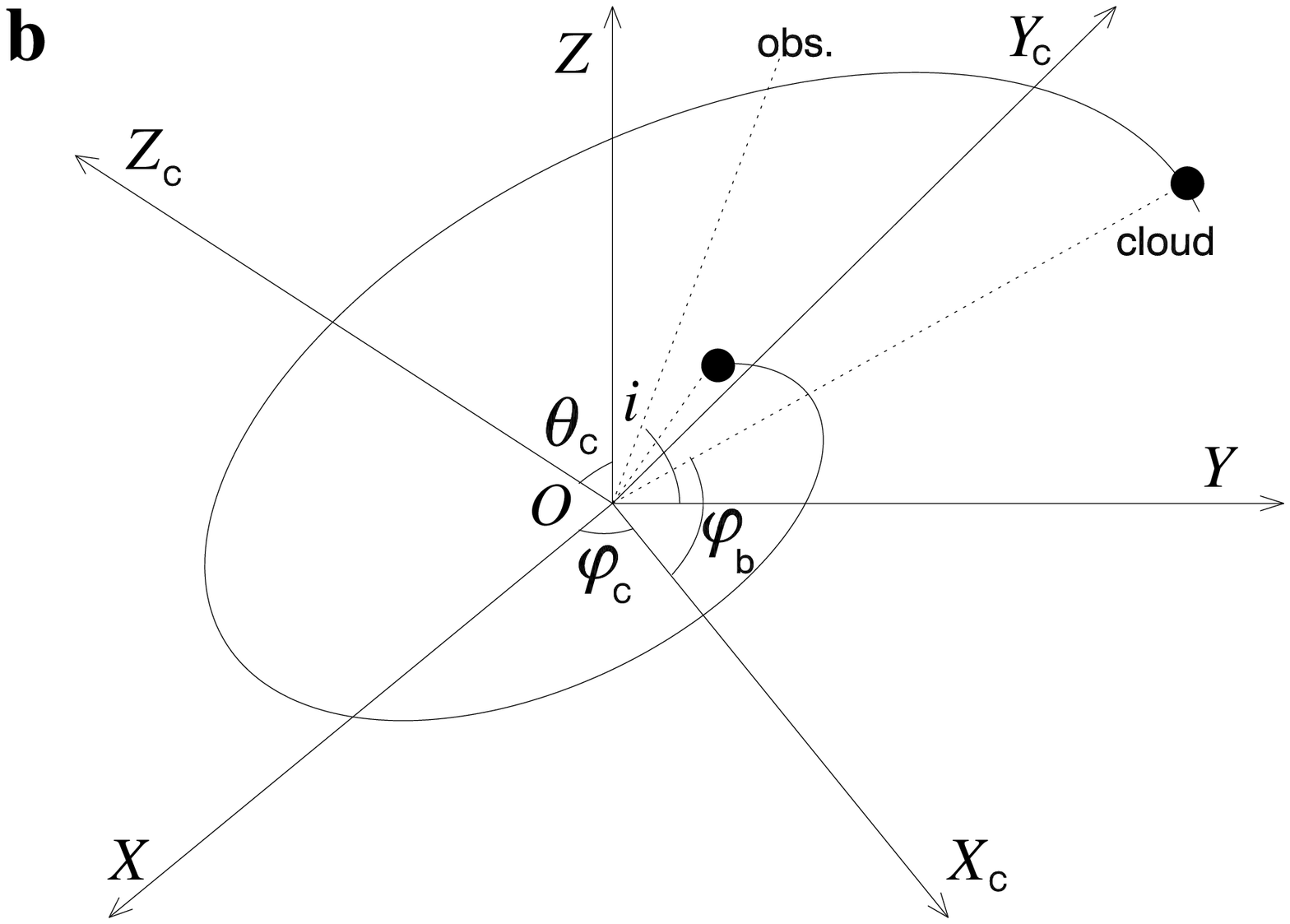}
\caption{\footnotesize {\bf The scheme of the present model.} Panel {\bf a} shows a 
cartoon of the tidal disruption of clumps moving to the inner region of the torus. 
The red and blue colours indicate redshifts and blue shifts of line emissions after
tidal disruption of clumps from the torus. Panel {\bf b} is the frame for calculations 
of profiles, observers are in $YOZ$-plane of the $XYZ$-frame. One of 
the clumps is orbiting in the $\XC O\YC$-plane, which is obtained by rotating the $X$-axis 
around the $Z$-axis ($\varphiC$) and rotating the $OZ$-axis around the $O\XC$-axis ($\thetaC$).  
The $O\YC$-axis is the major axis of the elliptical orbit and the $O\ZC$-axis is the normal 
direction of the orbital plane. Capture planes are defined as the orbital planes of individual 
clumps. $\Theta_{\rm torus}$ is the torus angle, which is defined by the angle from the mid 
plane to the inner edge of the torus (from $OY$-axis to the edge in the $YOZ$-plane).
}
\end{figure}

\newpage
\begin{figure}
\centering
\includegraphics[angle=0,origin=c,trim=100pt 90pt 10pt 50pt, width=1.1\textwidth]{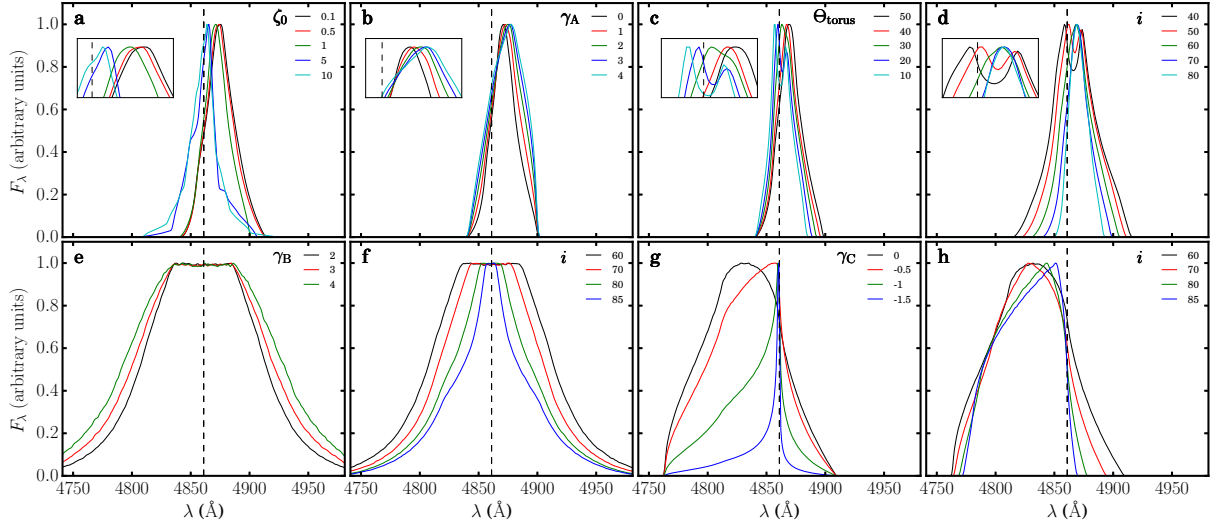}
\vglue 0.85cm
\caption{\footnotesize {\bf Various profiles of the models with a wide ranges of parameters.}
Panels {\it a-d} show spectra for type A clouds, {\it e-f} for the type B and {\it g-h} for the 
type C. The small boxes are zoom-in profiles of the peaks to clarify their fine structures.
In panels {\it a-d}, we show the profile dependence on ($\zeta_0, \gammaA, \Theta_{\rm torus},i$),
respectively, for given other parameters with typical values of 
$(\zeta_0,\gammaA,\Theta_{\rm torus},i$)=($1, 0,60^{\circ},70^{\circ}$), $\xi_{\rm A}=0.9$
and $\gamma=0$.
Type B clouds have symmetric profiles, but their widths change with cloud spatial distributions and 
inclinations. In panel {\it e}, $\gammaB=(2,3,4)$ whereas dependence on
$i$ and $(\gamma,\gammaB)=(0,2)$ is shown in panel {\it f}.
Type C clouds ejected by tidal dynamics generally show profiles with blue shifts. In panel {\it g},
the profile dependence on $\gammaC$ is shown for a fixed $(\xi_{\rm C},i)=(2,60^{\circ})$ whereas the
dependence on $i$ for a fixed $(\xi_{\rm C},\gammaC)=(2,0)$ in panel {\it i}. The dashed lines indicate 
$\lambda_0=4861$\AA.
}
\end{figure}

\newpage

\begin{figure}
\centering
\includegraphics[angle=0,origin=c,trim=120pt 90pt 10pt 80pt, width=1.11\textwidth]{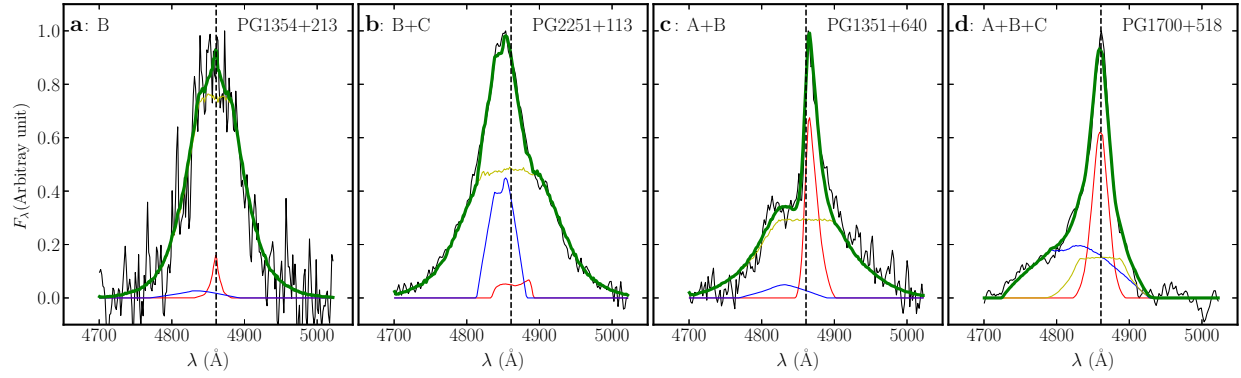}
\vglue 1cm
\caption{\footnotesize 
{\bf Best-fit models characterising the four different types of H$\beta$ profiles.} 
Object names and types are indicated in the panels. Black lines are observed spectra,
the red line is from type A clouds, the yellow from
type B, and the blue from type C, and green is the total. The dashed lines indicate
$\lambda_0=4861$\AA. We have $\chi^2=(1.6,2.2,2.6,3.1)$ for the best fittings of 
object $a-d$, respectively. The model parameters are given in Supplementary Table 2.}
\end{figure}

\begin{figure}
\centering
\includegraphics[angle=0,origin=c,trim=70pt 90pt -20pt 80pt, width=0.42\textwidth]{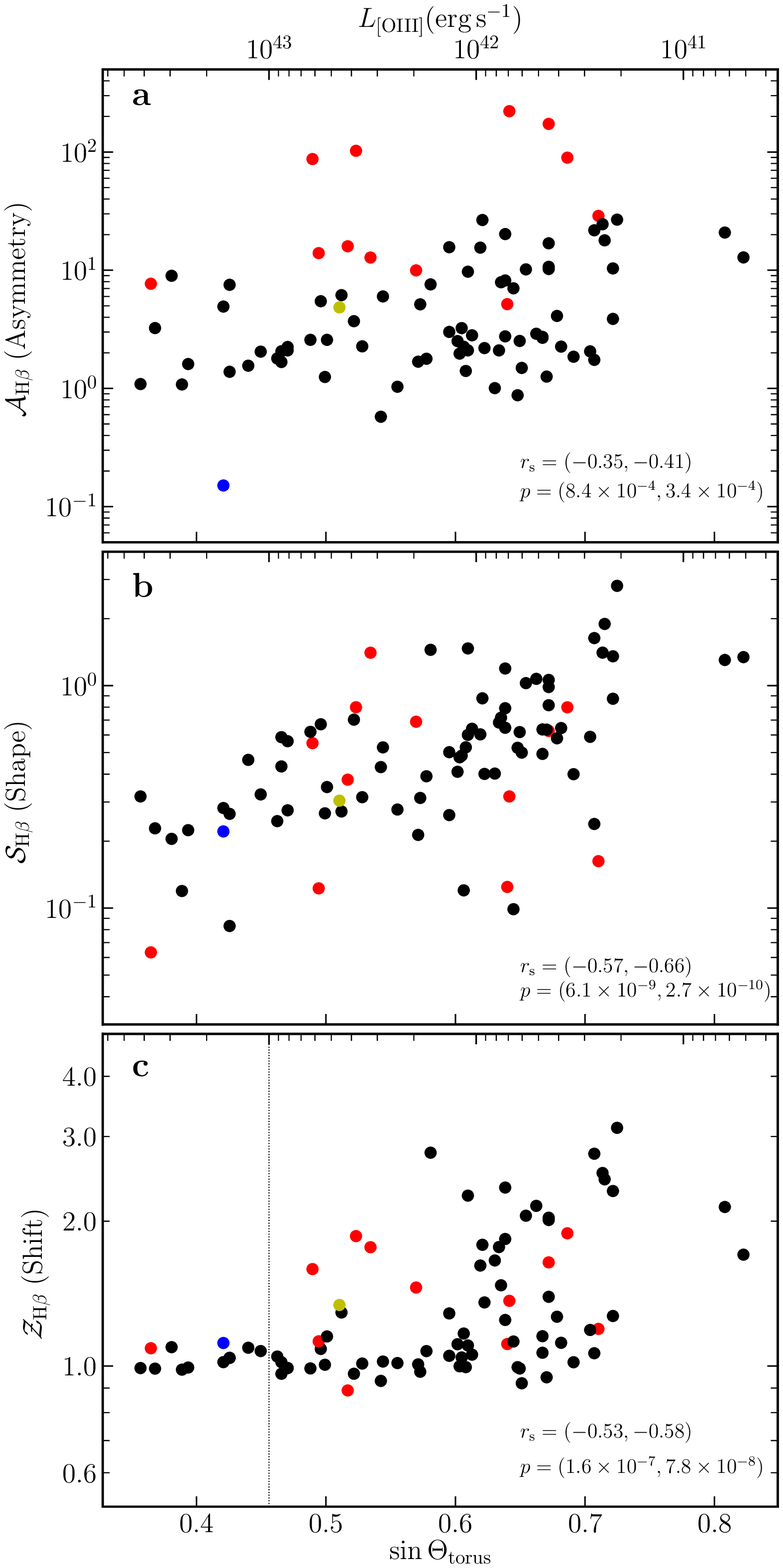}
\vglue -1.0cm
\caption{\footnotesize 
{\bf Correlations of asymmetries, shapes and shifts with \oiii\, luminosity as a proxy of torus 
angles} (the $\Theta_{\rm torus}-\Loiii$ relation is given by Equation \ref{Theta}). The correlations 
are among the most significant in PG quasars\cite{Boroson1992}. The red points are (A+B) objects, the 
one blue point and one yellow are (B+C) and B objects, respectively. Spearman correlation coefficients 
and null-probabilities are indicated on the panels (the first values are for the entire sample and 
the second for all black points). The dependence of $\calA$, $\calS$ and $\calZ$ on 
$\Theta_{\rm torus}$ indicates the origin of the BLR from torus. Spectral shifts increase only in 
quasars with $\Loiii\lesssim 10^{43}\,\ergs$ (the dotted line). Average error bars of  $\calA$, $\calS$ 
and $\calZ$ are discussed in the Methods. 
}
\end{figure}

\newpage


\begin{methods}

\subsection{Comparison with the Galactic center}
Recently the black hole at the Galactic center was observed to capture the molecular cloud 
G2, massing $\sim 3M_{\oplus}$, at a distance of $\sim 2400$ Schwartzschild 
radii\cite{Gillessen2012,Pfuhl2015}. The tidal interaction of the black hole and G2 produced 
a number of observational features\cite{Pfuhl2015,McCourt2015}. 
As we showed for typical values of parameters, the range of captured mass nicely 
spans the properties of the G2 cloud in the Galactic center. Such an event motivated us to 
think about other galaxies including AGNs: tidal captures of molecular clouds could be a 
common phenomena in galactic centers. As a reservoir of gas, the dusty torus provides a 
potential source of BLR gas via tidal captures, which could unite the torus, BLR, and 
accretion discs in AGNs. 

The mass of captured clouds is sensitive to black hole mass, very sensitive to the ratio 
of $\epsilon$, and also sensitive in particular to the dust sublimation temperature. Our 
estimate is only valid for AGNs with normal Eddington ratios since the sublimation radius 
depends on the spectral energy distributions (SEDs) (SDSS quasars have 
$L_{\rm Bol}/L_{\rm Edd}=0.1\sim 0.5$ distributed in a 
fairly narrow range, see Figure 12 in Ref.\cite{Shen2008}). 
Considering a wide range of black hole mass, $M_{\rm C}$ is mostly controlled by $\bhm$ and 
$T_{\rm sub}$. We note that $T_{\rm sub}$ can reach even 2000\,K influenced by effects of 
gas density or on chemical composition\cite{Laor1993}, leading to a vast change of 
$M_{\rm C}$ by a factor of 90 due to the extreme sensitivity of 
$M_{\rm C}\propto T_{1500}^{-15.6}$. This strong dependence implies that the BLR clouds 
rely on dust composition (which is related to star formation 
history of accretion discs\cite{Collin1999,Wang2012}, circumnuclear regions, and even of bulge 
regions). This could explain the dependence of metallicity on black hole mass or Eddington 
ratio in light of sensitivity of line flux ratios to 
metallicity\cite{Hamann2002,Warner2003,Matsuoka2011}. 
Further exploring these relations will aid our understanding of 
issues of galactic center evolution. We keep all of these as open hypotheses in mind, which 
are needed to explore in detail the capture of clumps through numerical simulations.

\subsection{Dynamics of tidally disrupted clumps.}
A dusty torus is recognized today to exist in AGNs generally. There is increasing evidence 
that the torus is not uniform, but rather composed of randomly moving dusty clumps; for 
instance, interferometric mid-infrared observations of NGC 1068 show the presence of dusty 
clumps\cite{Jaffe2004}. Clumpy torus models are now commonly employed to explain the 
near-to-mid infrared continuum in quasars and Seyfert 
galaxies\cite{Nenkova2008,Nenkova2002,Mor2009,Mor2012,Lira2013,Netzer2015,
Ichikawa2015,Fuller2016,Hern2016,Audibert2017}. The discovery of radial motion
of ionised gas deduced from quasar spectra\cite{Gaskell1982}, and in particular, the 
systematical shifts of the intermediate component of H$\beta$ line with the same shifts of 
\feii\, in SDSS large sample of quasars, imply a likely supply of the BLR gas from the inner 
edge of torus\cite{Hu2008a,Hu2008b}. 

The tidally captured clumps are undergoing 
complicated processes. It has been shown that collisions among clumps inside the torus 
are an efficient way to lose their angular momentum, leading to an inflow to fuel the central 
black hole. The rate of mass inflow depends on several factors, such as:  
the inelasticity of collisions ($\delta_{\rm diss}$), the fraction of the cloud's mass which 
participates in the collisions ($f_m$), the covering factor of the torus (${\mathscr{C}}$), the 
dispersion velocity ($\Delta v$), and the column density of clumps ($N_{\rm H,t}$). The rates 
can be written as (see Equation 15 in Ref.\cite{Krolik1988})
\begin{equation}\label{mdot}
\dot{M}\approx 1.0\, R_{\rm 0.4}^{1/2}M_8^{1/2}\left(\frac{f_m}{0.2}\right)
                   \left(\frac{\delta_{\rm diss}}{0.3}\right)
                   \left(\frac{{\mathscr{C}}N_{\rm H,t}}{10^{24}\rm cm^{-2}}\right)
                   \left(\frac{\Delta v_z}{v_{\rm orb}}\right)
                   \left(\frac{\Delta v}{v_{\rm orb}}\right) 
                   \sunm\,{\rm yr^{-1}},
\end{equation}
where $R_{0.4}$ is the inner edge of the torus in units of 0.4 parsec, $v_{\rm orb}$ is the 
orbital velocity at $R_{0.4}$, $\Delta v_z$ and $\Delta v$ are the dispersion velocities of 
clumps in the vertical and total directions, respectively. Details of the colliding clumps,
such as $f_m$ and $\delta_{\rm diss}$ can be found in Ref.\,13. The ratios of 
$\Delta v/v_{\rm orb}$ 
and $\Delta v_z/v_{\rm orb}$ determine thickness of the torus. Actually, collisions of clumps 
provide a physical mechanism to maintain the thickness of the tori for AGN unification schemes. 
Given the internal physics of clumps ($f_m$, $\delta_{\rm diss}$ and $N_{\rm H,t}$), on the 
other hand, $\mathscr{C}$ also plays a key role in the mass rate of the inflow. Equation
(\ref{mdot}) shows an evolutionary rate of mass inflow with geometry and covering factors of
tori\cite{Muller2013}. The improved model of clumpy accretion discs\cite{Vollmer2002} provides 
a similar rate of mass inflow and detailed investigations of numerical simulations are needed 
for torus\cite{Hopkins2012}.

The G2 cloud has been investigated by numerical simulations that include ablation, 
evaporation and compression, the complex dynamics incorporating drag 
forces that are either linear or proportional to the square of the velocity\cite{Pfuhl2015},
and radiation pressure (see a series of equations for details in Section 4 in 
Ref.\cite{Burkert2012}), however, for the captured clumps in AGNs, they 
are undergoing variations of thermal states due to cooling and heating of the 
accretion disc. Several other factors can be excluded by the following considerations. 
The heating power generated by the self-gravity of the clumps is of order
$\dot{E}_{\rm sg}\approx GM_{\rm C}^2\dot{R}_{\rm C}/R_{\rm C}^2
                 \approx 1.4\times 10^{28}\,M_{\rm C,2.4}^2R_{14}^{-2}\dot{R}_{7}\,\ergs$, 
where $R_{14}=R_{\rm C}/10^{14}$cm is the typical size of the clumps, 
$\dot{R}_{\rm C}=dR_{\rm C}/dt$ is the contracting velocity, 
$\dot{R}_{7}=\dot{R}_{\rm C}/10^7{\rm cm~s^{-1}}$.
The compression power depends on the density and temperature of the surrounding medium, 
which usually scales as $(n_{\rm f},T_{\rm f})\propto R^{-1}$ in the Galactic 
center\cite{Xu2006,Schartmann2012}. Following this relation, we take 
$T_{\rm f}=4.1\times 10^7\,D_{\rm sub,0.4}^{-1}$K and 
$n_{\rm f}=3.2\times 10^3\,D_{\rm sub,0.4}^{-1}{\rm cm^{-3}}$, the compression power is 
given by $\dot{W}=4\pi R_c^2P_{\rm f}\dot{R}_{\rm C}
\approx 2.3\times10^{31}D_{\rm sub,0.4}^{-2}R_{14}^{2}\dot{R}_{7}\ergs$,
where $P_{\rm f}=n_{\rm f}kT_{\rm f}$ is the ram pressure, $k$ is the Boltzmann constant,
and $D_{\rm sub,0.4}=D_{\rm sub}/0.4$pc.
The ratio of AGN heating to the compression is about 
$L_{\rm AGN}D_{\rm sub}^{-2}/\dot{W}R_{\rm C}^{-2}\approx 3\times 10^5$
for $L_{\rm AGN}\sim 10^{45}\ergs$,
indicating that heating fluxes of compression and self-gravity are much smaller than that of 
the accretion disc in AGNs, but it may be very important for G2 in the Galactic center. A
thermal balance is then expected between the heating of AGN radiation and cooling of captured
clumps.

The ablation timescale is sensitive to the density ratio of the clouds
to their surroundings, which is approximately given by
$t_{\rm abl}\approx 10^8\,R_{14}V_3^{-1}\left(n_{\rm C}/10^7n_{\rm f}\right){\rm yrs}
=10^8\,R_{14}V_3^{-1}n_{10}D_{\rm sub,0.4}\,$yrs (Equation 6 in Ref.\cite{Burkert2012}), 
where $n_{10}=n_{\rm C}/10^{10}{\rm cm^{-3}}$, $V_3=V_{\rm C}/10^3\kms$ is the clump
velocity relative to their surroundings and $n_{\rm f}$ is used. Here we keep the assumption 
that clouds hold equilibrium with their surroundings. We find that $t_{\rm abl}$
is much longer than the free-fall timescale of 
$t_{\rm ff}=D_{\rm sub}/V_{\rm ff}\approx 375\,\epsilon^{3/4}M_8^{1/4}T_{1500}^{-3.9}\,$yrs.
The evaporation timescale, driven by thermal conduction of the hot surrounding medium, is 
$t_{\rm evap}\approx 883\,\epsilon^{9/4}M_8^{-3/4}N_{24}T_{1500}^{-3.9}\,$yrs for the saturated 
evaporation (Equation 64 in Ref.\cite{Cowie1977}). Considering uncertainties 
of the parameters, we may have $t_{\rm evap}\sim t_{\rm ff}\ll t_{\rm abl}$, indicating that 
the ablation can be completely neglected, but the evaporation is important,
in particular, it may govern differences of high/low ionisation line regions formed
during the spiral-in of the captured clumps. 

A complete calculation of all the physical details of the captured clumps should invoke all 
of equations of the dynamics, thermodynamics, and radiation. We need to incorporate the equations 
listed in Ref.\cite{Burkert2012} involving the cooling and heating of the accretion discs along 
with the volume compression driven by the ram pressure. These processes invoke thermal 
instability\cite{Mathews1987}, leading to the rapid formation of BLR clouds (with the typical density 
of $\sim 10^{10-11}{\rm cm^{-3}}$) from the tidally captured clumps during the cooling timescale 
of the clumps of $t_{\rm cooling}\approx 14\,T_4\left(n_{10}\Lambda_{23}\right)^{-1}$s,
where $T_4=T/10^4$K and
$\Lambda_{23}=\Lambda/10^{-23}{\rm erg~s^{-1}cm^3}$ is the cooling function\cite{Sutherland1993}. 
Such a calculation will self-consistently generate dynamics and radiation of emission lines
from the clumps (to be carried out in an separate paper).
We also note that some clouds could be destroyed\cite{Mathews1986,Mathews1990}, however, the 
clumps are continually supplied by tidal captures from the torus.
In such a way, the BLR maintains a quasi-stationary state through 
balancing captures and mergers with accretion discs as done for the estimation of total clouds.

There are some simplified 
analytical approaches to the dynamics of clouds in the BLR, such as for the case of
constant mass, but they are likely insufficient\cite{Krause2011,Plewa2013,Shadmehri2015}.
While a detailed treatment of the dynamics of a tidally captured clump is beyond the scope of 
the present paper, we consider how the major characteristics of the above dynamical scenario 
will manifest observationally. In the frame of the $\XC\YC\ZC$-plane as the tidal disruption 
plane shown in Figure 1{\it b}, for type A clouds, we describe the simplified dynamics  as
\begin{equation}\label{va}
v_{_{\rm A}}=v_{_{\rm A}}^0\left(R/R_{\rm T}^0\right)^{-\alphaA},
\end{equation}
and angular velocity
\begin{equation}\label{omegaa}
\omega_{_{\rm A}}= \omega_{_{\rm A}}^0\left(R/R_{\rm T}^0\right)^{-\betaA},
\end{equation}
for $R>R_0$,
where $\omega_{_{\rm A}}=d\varphib/dt$ and 
$v_{_{\rm A}}=-dR/dt$. We assume that tidal disruption happens at a radius of $R_{\rm T}^0$ 
and an angle of $\varphi_0$ with $v_{_{\rm A}}^0$ and $\omega_{_{\rm A}}^0$. Here $R_0$ is the 
circularized radius due to drag forces. Type A and B cloud orbits are tangent at the pericentre 
point.

Combining Equations (\ref{va}) and (\ref{omegaa}) for $R>R_0$, we can write the trajectories of type A 
clouds
\begin{equation}\label{trajectory}
r=\left[r_{\rm torus}^{\Gamma}-\zeta_0^{-1}\Gamma
             \left(\varphi_{\rm b}-\varphi_0\right)\right]^{1/\Gamma},
\end{equation}
where $r=R/R_{\rm T}^0$, $\Gamma=(1+\alphaA-\betaA)$, 
$r_{\rm torus}=R_{\rm torus}/R_{\rm T}^0$ is the inner edge of the torus, and
$\zeta_0=R_{\rm T}^0\omega_{\rm A}^0/v_{\rm A}^0$ mainly determining the spiral-in trajectory. 
Equation (\ref{trajectory}) generates various kinds of spiral-in trajectories and describes the 
clouds moving under the influence of the drag forces. The maximum 
$\varphib$ for the spiral-in clouds is given by
$\varphi_{\rm max}=\varphi_0+\zeta_0\Gamma^{-1}
                  \left(r_{\rm torus}^{\Gamma}-r_0^{\Gamma}\right)$,
where $r_0=R_0/R_{\rm T}^0$. We note that $\Gamma$ is a variable, but it can not span a large 
range due to practical limits on $\alphaA$ and $\betaA$. The larger $\zeta_0$, the more cycles 
of the spiral-in. With Equation (\ref{trajectory}), we have velocities of
\begin{equation}\left(\begin{array}{l}
\vXC^{\rm A} \\
\vYC^{\rm A} \\
\vZC^{\rm A}\end{array}\!\!\!\right)
=-v_{\rm A}^0\left(\begin{array}{c}
    r^{-\alphaA}\cos\varphib+\zeta_0r^{1-\betaA}\sin\varphib \\
    r^{-\alphaA}\sin\varphib-\zeta_0r^{1-\betaA}\cos\varphib \\
    0\end{array}\right).                    
\end{equation}
and the total velocity of 
\begin{equation}
v_{\rm A}=v_{\rm A}^0\left[r^{-2\alphaA}+\zeta_0^2r^{2(1-\betaA)}\right]^{1/2},
\end{equation}
in the $\XC\YC\ZC$-plane. The mass distribution of the clouds is assumed to be 
$N_{\rm A}\propto v_{\rm A}^{\gammaA}$ owing to insufficiently understood dynamics. 
The index $\gammaA\gtrsim 0$ is expected for mass conservation.

Type B clouds orbit around the black hole,
but they gradually drift onto the accretion disc. For the calculation of
spectral profiles, this is equivalent to a series of rings with different orientations 
and radii. In the $\XC\YC\ZC$-frame, we have 
$\vXC^{\rm B}=-v_{\rm K}\sin \varphib$, $\vYC^{\rm B}=v_{\rm K}\cos \varphib$ and 
$\vZC^{\rm B}=0$, where $v_{\rm K}$ is the Keplerian velocity.
For different capture events, $R$ has a distribution from $R_0$ to $R_{\rm in}$,
where $R_{\rm in}$ is the inner edge of the BLR (we take $R_{\rm in}<\rblr$ and assume
clouds follow $N_{\rm B}\propto v_{\rm B}^{\gammaB}$). This component could contribute 
the majority of the observed broad emission. We note here that type A clouds 
observationally overlap the type B clouds if $\zeta_0$ is large enough.

Type C clouds also undergo interaction with their surroundings, but they escape from
the black hole potential. We simplify their trajectory as a straight line after the 
pericentre. This trajectory is simply given by
$R=R_{\rm T}^0/\cos\varphi_{\rm b}$,
in the $\XC\YC\ZC$-frame. With insufficient details of the tidal dynamics, we assume the 
approximate initial velocity
of the ejected clouds to be $v_{\rm C}^0=\xi_{\rm C} v_{\rm K}^0$ and $\xi_{\rm C}\gtrsim 1$,
where $v_{\rm K}^0$ is the Keplerian velocity at $R_{\rm T}^0$. For the decelerating dynamics
of the escaping clouds, we assume velocities in the form
$\vYC^{\rm C}=v_{\rm C}^0\left(R/R_{\rm T}^0\right)^{-\alphaC}$, $\vXC^{\rm C}=\vZC^{\rm C}=0$,
and the spatial distribution is $N_{\rm C}\propto v_{\rm C}^{\gammaC}$, where $\alphaC>0$
(deceleration by drags) or $\alpha_{\rm C}<0$ (acceleration driven by radiation pressure 
of accretion discs), $N_{\rm C}$ is the type C cloud number for unit radius, and $\gammaC$ is 
the power-law index. The acceleration driven by radiation pressure depends on several factors, 
including column density, metallicity, and the degree of ionisation of the clumps. 

\subsection{Ejection fraction.}
The fraction of type C clouds depends on details of their internal states and tidal interaction. 
We first consider the simpler case of a solar-type star with mass of $m_*$ and radius of $r_*$, 
with the side extending away from the black hole having an 
extra velocity above orbital, whereas the side closest to the black hole has a comparable 
velocity deficit. This black hole tidal torque therefore forces the star to rotate until 
disruption, when rotational energy is comparable with self-gravitational energy. 
Specific energies of the stellar gas are distributed in a range of 
$E_{\rm max}=\left(Gm_*/r_*\right)\left[\left(\bhm/m_*\right)^{1/3}-1\right]$ 
to $E_{\rm min}=-\left(Gm_*/r_*\right)\left[\left(\bhm/m_*\right)^{1/3}+1\right]$ 
in a single flyby\cite{Rees1988}. Taking a uniform distribution of energy over stellar 
mass\cite{Rees1988}, confirmed by numerical simulations\cite{Evans1989}, the ejection 
fraction can be obtained by 
$f_{\rm C}=E_{\rm max}/\left(E_{\rm max}-E_{\rm min}\right)\approx0.5$ due to 
$\left(\bhm/m_*\right)^{1/3}\gg1$. This estimate is valid only for solar-type stars since 
the dynamical timescale for a captured star holds $t_{\rm dyn}\ll t_{\rm orb}$, where 
$t_{\rm orb}$ is the orbital timescale. However, it does not directly apply to the clumps 
we consider. If $t_{\rm dyn}\gg t_{\rm orb}$ holds, the clumps will be stretched into a 
ring. When $t_{\rm dyn}\gtrsim \eta_0 t_{\rm orb}$ holds, where $\eta_0$ is a significant 
fraction, the clumps will be partially ejected. 

It is difficult to derive expression of $f_{\rm C}$ without details of the internet states 
of the clumps, but we can derive its dependence on ($\epsilon,\bhm,T_{\rm sub})$ following 
the above arguments. The mass of the tidally captured clumps governed by the sound wave, 
and is given by 
$M_{\rm C}^{\prime}\approx \frac{4\pi}{3}n_{\rm C}m_{\rm p}\left(c_{\rm s}t_{\rm orb}\right)^3$ 
during one orbit $t_{\rm orb}$, where $c_{\rm s}$ is the sound speed.
Actually, $M_{\rm C}^{\prime}$ is the effective mass as 
part of one clump can be tidally disrupted like a star, and half its mass
ejected by the tidal force. We thus have the fraction of 
$f_{\rm C}\approx 0.5 M_{\rm C}^{\prime}/M_{\rm C}\propto \left(t_{\rm orb}/t_{\rm dyn}\right)^3$, 
where $t_{\rm dyn}$ is the dynamical timescale and 
$M_{\rm C}=4\pi n_{\rm C}m_{\rm p} R_{\rm C}^3/3$. The self-gravity free-fall timescale is  
$t_{\rm sg}=\left(4\pi Gn_{\rm C}m_{\rm p}/3\right)^{-1/2}=1.57\times 10^{11}\,T_{1500}^{1/2}D_{\rm sub,0.4}$, 
where we use the pressure balance 
$n_{\rm C}T_{\rm C}=n_{\rm f}T_{\rm f}\approx 1.3\times 10^{11}D_{\rm sub,0.4}^{-2}$, where
$T_{\rm C}$ is the temperature of clouds.
Setting $t_{\rm dyn}=t_{\rm sg}$ and $T_{\rm C}\approx T_{\rm sub}$, we have 
$t_{\rm orb}/t_{\rm dyn}\approx 0.48\,M_8^{-1/2}D_{\rm sub,0.4}^{1/2}T_{1500}^{-1/2}$,
yielding 
\begin{equation}\label{fc}
f_{\rm C}\approx 0.06\,\epsilon^{3/4}M_8^{-3/4}T_{1500}^{-5.4}.
\end{equation}
Holding $T_{1500}$ constant, $f_{\rm C}$ increases with $\epsilon$ and decreases with $M_8$.
This factor $f_{\rm C}$ is determined by the competition between the self-gravity and tidal 
forces. Only the self-gravitating part of the clumps can be tidally disrupted and the rest  
is tidally captured and is not able to adjust through self-gravity and tidal 
forces, in order to form type A and B clouds. Since $\epsilon$ represents Eddington ratios 
(or accretion rates), Equation (\ref{fc}) shows that AGNs with high accretion rates and 
less massive black holes will produce more type C clouds,
namely, more outflows. This is generally consistent with narrow line Seyfert 1 galaxies 
observed in UV band\cite{Leighly2004}, which are generally regarded as AGNs containing less 
massive black holes with high accretion rates. This estimate shows type C clouds only 
contribute a small fraction to the total profiles. We should keep in mind that the 
current estimation of $f_{\rm C}$ is obtained by a highly simplified treatment.

\subsection{Emission-line profiles.} 
Using the rotation matrix, we transform cloud velocities in the $\XC\YC\ZC$ into the $XYZ$-frame 
for the projected velocity to the line of sight
\begin{equation}\left(\!\!\begin{array}{l}
\vlos_{_{\rm A}} \\
\vlos_{_{\rm B}} \\
\vlos_{_{\rm C}}\end{array}\!\!\right)
=\left(\!\!\begin{array}{c}
-v_{\rm A}^0\left(q_1\cos\varphib+q_2\sin\varphib\right)\\
v_{\rm K}(-p_1\sin\varphib+p_2\cos\varphib)\\
v_{\rm C}^0p_2\left(\cos\varphib\right)^{\alphaC}\\
\end{array}\!\!\right),                    
\end{equation}
for type A, B and C clouds, where 
$q_1=p_1r^{-\alphaA}-p_2\zeta_0r^{1-\betaA}$,
$q_2=p_1\zeta_0r^{1-\betaA}+p_2r^{-\alphaA}$,
$p_1=\sin\varphiC\cos i$ and $p_2=\sin\thetaC\sin i+\cos\varphiC\cos\thetaC\cos i$.
For extremely high $\dot{\cal R}$, the global distribution of captured clumps has a stationary 
configuration. This allows us to treat the problem in a stationary way. Taking a narrow 
Gaussian as our emission-line profile for individual clouds, we can determine the emission 
from the type A, B, and C components for an individual tidal capture:
\begin{equation}\label{flambda}
f_{\lambda}^{\rm A}=\int_{\varphi_0}^{\varphi_{\rm max}}N_{\rm A}(v_{_{\rm A}})
             \mathbb{G}(\lambda)d\varphib,
\end{equation}
\begin{equation}
f_{\lambda}^{\rm B}=\int_{r_0}^{r_{\rm in}}rdr
             \int_{0}^{2\pi}N_{\rm B}(v_{_{\rm B}})\mathbb{G}(\lambda)d\varphib,
\end{equation}
and
\begin{equation}
f_{\lambda}^{\rm C}=\int_{0}^{\pi/2}N_{\rm C}(v_{_{\rm C}})\mathbb{G}(\lambda)d\varphib,
\end{equation}
where $\mathbb{G}(\lambda)=\left(\sqrt{2\pi}\Delta \lambda\right)^{-1}
\exp\left[-(\lambda-\lambda_0)^2/2(\Delta\lambda)^2\right]$ is the Gaussian function and 
$\lambda_0$ is the intrinsic wavelength and $\Delta\lambda$ is the width of the line.
Assuming high capture rates with random directions, we can write the probability distribution 
of orientations as proportional to $\left(\cos\theta_{\rm C}\right)^{\gamma}$, where 
$\gamma$ depends on the vertical structure of the torus. We now write the emission from 
successive captured clumps from the upper mid-plane
\begin{equation}\label{Flambda}
F_{\lambda}= \int_0^{2\pi}d\varphi\int_{0}^{\Theta_{\rm torus}}d\theta
                   \left(\cos\theta_{\rm C}\right)^{\gamma}
                   \left(f_{\lambda}^{\rm A}+f_{\lambda}^{\rm B}+f_{\lambda}^{\rm C}\right).
\end{equation}
The fractions from each components can be obtained from
\begin{equation}
f_{\rm A}=\frac{F_{\rm A}}{F_{\rm tot}};~~~
f_{\rm B}=\frac{F_{\rm B}}{F_{\rm tot}},
\end{equation}
where $f_{\rm C}=1-f_{\rm A}-f_{\rm B}$, $F_{\rm A,B,C}$ are the fluxes from type A, B, C 
clouds, respectively, and $F_{\rm tot}$ is the total flux.
Here we include only captures from above the mid-plane of the torus, assuming that 
the disc is optically thick. In Equations (\ref{flambda}) and (\ref{Flambda}), we assume that 
clouds are optically thin and radiating H$\beta$ photons isotropically.

Supplementary Table 1 summarises all the parameters and ranges used in our model. 
The first ten parameters are the primary drivers of the observed profiles whereas others are 
auxiliary and can be constrained by either observations or determined to reside within small 
ranges by physical constraints.
We take $R_{\rm torus}/\Rg=8.5\times 10^4\,L_{45}^{1/2}T_{1500}^{-2.6}M_8^{-1}$ 
as the outer boundary of tidal captures consistent with the 
results from NIR-RM\cite{Koshida2014}. The circularized radius of disrupted
clumps can be conveniently approximated by the reverberation mapping relation of  
$R_0\gtrsim R_{\rm BLR}$. We lack the details of specific tidal disruption locations
without internal states of the clumps, but do know that $R_{\rm T}^0\in [R_0,R_{\rm torus}]$ 
holds. For an illustration of the present model, we assume $R_{\rm T}^0/\Rg=10^4$ and 
$\varphi_0=\pi/2$. Assuming the total initial velocity of 
$V_{\rm tot}^0=\xi_{\rm A}v_{\rm K}$ at $R_{\rm T}^0$, 
we have $v_{\rm A}^0=\xi_{\rm A}(\zeta_0^2+1)^{-1/2}v_{\rm K}$ from
$V_{\rm tot}^0=\left[\left(v_{\rm A}^0\right)^2+
                 \left(R_{\rm T}^0\omega_{\rm A}^0\right)^2\right]^{1/2}
               =(\zeta_0^2+1)^{1/2}v_{\rm A}^0$, 
where $\xi_{\rm A}$ is a constant ($\xi_{\rm A}<1$ 
for type A clouds since they are bound). Since type A clouds are bound, 
$\alphaA<1/2$ follows from $v_{\rm A}^0\le \left(G\bhm/R_0\right)^{1/2}$ as an infall
velocity less than the free-fall velocity and $\betaA>3/2$ for sub-Keplerian rotation. 
In our calculations, we take $\Gamma=-0.25$ ($\alphaA=0.45$ and
$\betaA=1.8$), $\xi_{\rm A}=0.9$ and $\gammaA=0$ for illustrations. For simplicity,
we assume that all the captured clumps are same and tidally disrupted at the same distance 
to the central black hole. This approximation is actually equivalent to 
a kind of averaged captured radii for the global system of captured clumps. Future improvement
of this approximation can be done by numerical simulations. We take 
$\lambda_0=4861$\AA\, and $\Delta\lambda=1$\AA\,  for the H$\beta$ line.

\subsection{Fitting H$\beta$ profiles of PG quasars}
We use a tested quasar spectral fitting procedure\cite{Hu2008a} to obtain H$\beta$ profile 
fits between ($4761-4961$)\AA\, of PG quasars after subtracting H$\beta$ narrow components, 
\feii, \oiii\, emission lines and continuum. We define 
$\chi^2=\sum_{i=1}^N\left(F_{\lambda_i}^{\rm mod}-F_{\lambda_i}^{\rm obs}\right)^2/\sigma_i^2$,
where $F_{\lambda_i}^{\rm mod}$, $F_{\lambda_i}^{\rm obs}$ and $\sigma_i$ are the
model, observed fluxes and errors, respectively, and $N$ is the number of wavelength points. 
For a given set of model parameters $\Pi$, their posterior probability distributions are 
given by the Bayes theorem
$p(\Pi|{\cal D})\propto \exp\left(-\chi^2/2\right)p(\Pi)$,
where $p(\Pi)$ is the prior distribution of model parameters. We assume it to be flat, namely, 
$p(\Pi)={\rm a~ constant}$. The Markov Chain Monte Carlo (MCMC) method is employed to find the 
best fit to data ${\cal D}$. 

As listed in Supplementary Table 1, there are 20 parameters for a full description of
the global model. We attempt to optimize fitting in two aspects: 1) to reduce number of 
the model parameters, and 2) to reduce degeneracy among the parameters. In order to achieve 
the two goals,  we choose the most sensitive parameters by taking into account their roles in 
the model. We note that emissivities should depend on radius  (expressed by some auxiliary 
parameters of $\gammaA$, $\gammaB$ and $\gammaC$), but we only choose
$\gammaA$ and $\gammaB$ ($\gammaC$ is regarded as one auxiliary because type C clouds are 
always small contribution to the total). For type A clouds, $\zeta_0$ is the 
major driver of spiral-in trajectories and infalling velocity shifts. 
In principle, $\xi_{\rm A}$ is independent of $\zeta_0$, but is degenerate with $\zeta_0$. 
We have three of ($\zeta_0,\xi_{\rm A},\gammaA$) for type A clouds. For type B 
clouds, $R_0$ (the inner and the outer radii of type A and B clouds, respectively)
connects type A and B clouds, $R_{\rm in}$ determines the width of the symmetric profiles, 
$f_{\rm B}$ does the relative flux to the total and $\gammaA$ determines the emissivity 
distributions of type B clouds. We have four parameters of ($R_0,R_{\rm in},\gammaB,f_{\rm B}$) 
for type B clouds.  For type C clouds, $\xi_{\rm C}$ determines the shifts of blue components of
spectra and $f_{\rm C}$ does the relative flux to the total. Inclination as the only external 
parameters of the model but still controls apparent profiles. All other parameters are listed
as auxiliary ones. Therefore, the majorities of parameters controlling the profiles are 
$\Pi=(\zeta_0,\gammaA,\xi_{\rm A},R_0,R_{\rm in},\gammaB,f_{\rm B}, \xi_{\rm C},f_{\rm C},i)$
and the auxiliary parameters are fixed numbers given in bold fonts in Supplementary Table 1. 

Torus angles play an important role in the integration of Equation (\ref{Flambda}). 
We used $\Theta_{\rm tours}$ as one of the fitting parameters, however, we 
found that the $(i,\Theta_{\rm torus})$-degeneracy 
is too significant to yield reliable results. Actually, this is the major degeneracy in the
model. Fortunately, the receding torus relation of AGN allows us to break the degeneracy. 
Assuming completely random orientations of AGNs, $\Theta_{\rm torus}$ can be estimated by 
$f_2=\Delta\Omega/2\pi=\sin \Theta_{\rm torus}$, where $f_2$ is the relative fraction of 
Type 2 AGNs\cite{Osterbrock1988,Tovmassian2001,Cao2005,Wang2005}. Torus angles follow 
from fractions of Type 2 AGNs statistically\cite{Reyes2008} 
\begin{equation}\label{Theta}
\sin\Theta_{\rm torus}=0.77-0.16\log L_{\rm [OIII],41},
\end{equation}
where $L_{\rm [OIII],41}=\Loiii/10^{41}\ergs$, which is consistent with other empirical relation
in Ref.\cite{Simpson2005,Maiolino2007}. We set a reasonable range
of $\Theta_{\rm torus}$ in the fittings through the scatters of Equation (\ref{Theta}).
Considering the average scatters of about  $\Delta f_2\approx 0.15$, we have 
$\Delta\Theta_{\rm torus}= \Delta f_2/\sqrt{1-f_2^2}\approx 10^{\circ}$, 
where $\langle f_2\rangle\approx 0.6$, as the average range of $\Theta_{\rm torus}$
for all objects.

Inclinations are difficult to estimate, but several efforts\cite{Marin2016} for quasars and
Seyfert 1 galaxies show $i=40^{\circ}\sim 90^{\circ}$ (orientations of PG quasars have not been 
systematically investigated). In our MCMC implementation, we include these physical constraints 
so as to alleviate some degeneracies of the $\Pi$-parameters. Moreover, we set a 
constraint of $i\ge\Theta_{\rm torus}$, which is a necessary condition for 
PG quasars as type 1 AGNs. We note that both $i$ and $R_{\rm in}$ have influences on  
profile widths and $f_{\rm B}$ determines its relative contribution to the total, leading to 
a degeneracy of $i$ with $f_{\rm B}$ and $R_{\rm in}$ in the fittings. This degeneracy 
can be alleviated by a self-consistent model discussed below.

Actually, it turns out that some of the $\Pi$-parameters are not critical (at least for individual 
objects) as shown by the resultant fittings. We would like to point out that the main goal of the 
PG quasar profile fitting exercise is to show the model characteristics by a comparison with data, 
rather than the detailed fitting fine structure of the profiles of any individual objects. This 
allows us to obtain the major statistical comparisons of the model with data. This can be done only 
after we build up a self-consistent model briefly discussed in ``Implications of the present model", 
in which numbers of some characterized parameters can be analysed.

All the fitted spectra are shown in Supplementary Figure 1 to illustrate the models for each of 
the quasars, 
with a few exceptions (see notes on them). Contributions of three cloud types to the total can 
be found by red, blue and yellow lines in the figure. The parameter $\Pi$-distributions are 
given in Supplementary Figure 2, showing that the parameters obtained by the fittings are reasonable.
We attempt to find potential relationships among the fitting parameters, but only find rather weak 
correlations. This indicates that the parameters likely have only weak degeneracies and thus 
the present options of fitting parameters are quite good. 

\subsection{H$\beta$ profiles in PG quasars}
With the decomposition of H$\beta$ profiles, we define the barycentre wavelength and flux as
$\bar{\lambda}=\int \lambda f_{\lambda}d\lambda/\int f_{\lambda}d\lambda$ and
$h=\frac{1}{2}\int f_{\lambda}^2d\lambda/\int f_{\lambda}d\lambda$
for the component $f_{\lambda}$, respectively. Supplementary Figure 3 illustrates a cartoon of the 
decomposed profiles for $(h,\bar{\lambda})$ as well as for the following three parameters. 
As a first-order approximation, the two parameters of 
$(h,\bar{\lambda})$ are complete to describe the profiles of each decomposed components 
(A and C). Asymmetries are mainly described by ratios of $h$ whereas shapes and shifts by 
$\Delta\lambda$.

Considering that component B is symmetric, we define the following parameters to characterize 
asymmetry, shape and shift of profiles
\begin{equation}\label{three-parameters}
\calA=\frac{h_{\rm A}}{h_{\rm C}},~~~
\calS=\frac{\Delta\lambda_{\rm C}+\Delta\lambda_{\rm A}}{\rm FWHM},
~~~{\rm and}~~~
\calZ=1+\frac{\Delta\lambda_{\rm C}-\Delta\lambda_{\rm A}}{\rm FWHM},
\end{equation}
where $\Delta\lambda_{\rm A}=\bar{\lambda}_{\rm A}-\lambda_0$, 
$\Delta\lambda_{\rm C}=\lambda_0-\bar{\lambda}_{\rm C}$, $\lambda_0=4861$\AA\,
and FWHM is full-width-half-maximum of the sum of broad components of the spectra.
{\cblue FWHM is a good indicator of line shape for Gaussian profiles, but not 
enough to describe general profiles. The physical meaning of the $\calS$ parameter 
indicates the relative width of one line with respect to its FWHM, representing 
the shapes of profiles. In Supplementary Figure 3, panel {\it b-d} show shapes with 
$\calS$. The parameter $\calZ$ describes the velocity differences between the type A 
and C clouds.} The Boroson and Green definitions\cite{Boroson1992} of similar parameters
are based on observed {\cblue total} profiles, but they are insufficiently physically
understood. However, the present $\calA$, $\calS$ and $\calZ$ are defined 
by components decomposed by the model, making it easy 
to explore the nature of profiles. We note that the shape and the shift defined by 
Ref.\cite{Boroson1992} don't correlate with $\Loiii$ (see their Table 3), but only the 
asymmetry. We find the present asymmetry correlates with ones defined 
by Ref.\cite{Boroson1992}.

{\cblue We can introduce $f_{\rm A}$ and $f_{\rm C}$ as
weights for 
\begin{equation}\label{three-parameters}
\calS^{\prime}=\frac{f_{\rm C}\Delta\lambda_{\rm C}+f_{\rm A}\Delta\lambda_{\rm A}}{\rm FWHM},
~~~{\rm and}~~~
\calZ^{\prime}=1+\frac{f_{\rm C}\Delta\lambda_{\rm C}-f_{\rm A}\Delta\lambda_{\rm A}}{\rm FWHM},
\end{equation}
which are parallel to Equation (15).
It is straightforward to show that $\calZ^{\prime}$ is equivalent to the definition
of mean shifts of the total profile with respect to $\lambda_0$.
From the barycenter wavelength of the total spectra given
by $\bar{\lambda}=\int \lambda f_{\lambda}d\lambda/F_{\rm tot}$, we have
$\bar{\lambda}=f_{\rm A}\bar{\lambda}_{\rm A}+f_{\rm C}\bar{\lambda}_{\rm C}+f_{\rm B}\lambda_0
              =f_{\rm A}\Delta\lambda_{\rm A}-f_{\rm C}\Delta\lambda_{\rm C}+\lambda_0$,
where $\bar{\lambda}_{\rm B}=\lambda_0$ is used because of the symmetry of component B. Denoting 
$\Delta\lambda=\bar{\lambda}-\lambda_0$, we have 
$\Delta\lambda=f_{\rm A}\Delta\lambda_{\rm A}-f_{\rm C}\Delta\lambda_{\rm C}$, leading to
relations of
$\calZ^{\prime}=1-\Delta\lambda/{\rm FWHM}$ and 
$\calS^{\prime}=(2f_{\rm A}\Delta\lambda_{\rm A}-\Delta\lambda)/{\rm FWHM}$. This demonstrates 
that $\calZ^{\prime}$ is independent of decompositions, but $\calS^{\prime}$ is still dependent. 
Actually, $\calZ^{\prime}$ is different from the Boroson-Green's definition of centroid at 3/4 
maximum from the rest wavelength in units of the FWHM. We plot $(\calS^{\prime},\calZ^{\prime})$ 
versus $L_{\rm [OIII]}$ in Supplementary Figure 4. Comparing with Figure 4, we find that the
($\calS^{\prime},\calZ^{\prime})- L_{\rm [OIII]}$ correlations get weaker,
but are still consistent generally. This suggests that the decomposition of total profiles 
may provide more physical insight.}

The error bars of $\calA$, $\calS$ and $\calZ$ are mainly determined by $\left(\Delta f/f\right)_{\rm A,C}$.
Using the averaged values of $\langle f\rangle_{\rm A,C}=(0.5,0.1)$ and 
$\left(\Delta f/f\right)_{\rm A,C}\approx (0.2,0.6)$ 
from Table 1, we have uncertainties of $\Delta\log(\calA,\calS,\calZ)\approx (0.27,0.14,0.14)$
on average. {\cblue ($\calS^{\prime},\calZ^{\prime})$ have the similar error bars.}
Here we use 10\% as the general uncertainty of FWHM.

\subsection{Inclinations of PG quasars}
Inclination angle is an important parameter, but extremely difficult to estimate.
The average inclination is $\langle i\rangle\approx 63^{\circ}$ obtained from the model fits to 
the PG sample. It is challenging to compare in a meaningful way the inclinations of individual 
objects with inclinations estimated by other methods, but the distributions of the presently 
determined inclinations are consistent with those of type 1 AGNs\cite{Marin2016}. Our analyses 
find no significant difference of BLR inclinations between radio-loud and radio-quiet quasars. 
This is interesting in light of the  comparison of our inclinations of radio-loud PG quasars to 
those estimated by the self-synchrotron Compton limit in radio-loud AGNs and quasars (see Figure 
7 in Ref.\cite{Ghisellini1993}), which are generally consistent. Moreover, the spectral 
slope in the $(1700-3000)$\AA\, range could be an indicator of inclination in broad absorption 
line quasars\cite{Baskin2013}. This motivates us to compare the slopes of
the optical-ultraviolet continuum of PG quasars\cite{Baskin2005} with the present inclinations. 
Here the slope is defined by $f_{\nu}\propto \nu^{-\alpha_{_{\rm OUV}}}$ measured between 
$(1549-4861)$\AA.  As shown by Supplementary Figure 5, the correlation is weak, but the trend 
is clear (it should be noted that $\alpha_{_{\rm OUV}}$ is defined in a significantly wider 
range than the slopes in the $(1700-3000)$\AA\, range, which might weaken the correlation). 
This lends support to the current inclinations from the present model. Similar 
tests can be done for a large sample of SDSS quasars\cite{Krawczyk2015} in the future.

\subsection{Other models}
There are a number of ideas about the origin of the BLR gas that have been investigated.
Most involve winds\cite{Shlosman1985,Murray1995,Proga2004}, failed dusty 
winds\cite{Czerny2011,Baskin2014} or magnetohydrodynamical (MHD) winds\cite{Emmering1992,Konigl1994} 
from accretion discs, radiation-pressure driven dusty outflows\cite{Gaskell2017},
discrete clouds embedded in a hot medium\cite{Krolik1981},
phenomenological models\cite{Collin1988,Xue1994,Eracleous1995,Goad2012} (see Table 4 of 
Ref.\cite{Wang2012} for a brief summary, or a recent paper\cite{Elvis2017}),  
and finally condensation of warm absorbers suggested very recently\cite{Elvis2017}. 
Disc wind models are problematic as they fail to meet observational constraints from 
velocity-resolved reverberation mapping of AGNs\cite{Gaskell2016}, to be inconsistent with 
evidence for the virialization relation of ${\rm FWHM}\propto \tau_{\rm H\beta}^{-1/2}$ in a couple 
of mapped AGNs (3C 390.3, NGC 3783, NGC 5548 and NGC 7469)\cite{Peterson2011}, where 
$\tau_{\rm H\beta}$ is the observed lags of H$\beta$ line,
or suffer from predictions of small covering factors of the BLR\cite{Murray1998}. MHD
models are quite flexible, but difficult to observationally test. 
X-ray eclipses by material with BLR properties indicate that 
discrete clouds are present in the BLR\cite{Risaliti2007} .
Many phenomenological models involve two  regions for low- and high-ionisation broad
emission lines separately (see the cartoon of Figure 1 in Ref.\cite{Collin1988}). The necessary 
conditions are extensively studied for broad emission lines: low-ionisation regions are
the outer part of accretion discs, but high-ionisation line regions with a spherical geometry
composed of discrete clouds whose infalling kinematics remains completely unknown (seemingly 
independent of accretion discs). Profiles have not been compared with observations, yet.
Recently, the failed dusty wind model, which is attractive to explain the observed $R-L$ relation, 
receives much attention. However, much work remains to be done in order to compare this 
model with observations, such as the calculation of profiles of emission lines from the winds, 
the dependence of dynamics on vertical structure of the accretion discs. Moreover, how to form \civ\, 
regions as high-ionisation regions (depending on radiation pressure\cite{Baskin2014} somehow)
which may be smaller by a factor of $5-10$ than H$\beta$ regions\cite{Kaspi2007}, fully remain 
open questions.

Outflows are apparently common among AGNs\cite{Crenshaw2003,Proga2000,Tombesi2014}, and, if 
they originate from the inner parts of accretion discs, there arises an interesting question: 
are the outflows in some way associated with the existing BLR clouds? In principle, they 
are spatially overlapped somehow if the outflows are approximately parallel to the disc surface, 
but the velocity-resolved reverberation mapping of AGNs through
spectroscopically monitoring campaigns shows evidence for inflows and rotation in most 
AGNs\cite{Denney2009,Grier2013,Du2016b}, and only two AGNs (NGC 3227 and Mrk 142) show evidence 
for outflows. This probably means that outflows are not the primary origin of the BLR, but 
also that strong interactions are not so common. However, more campaigns are needed to generally 
explore if there is such an interaction even if an outflow is not the primary source 
of the BLR. 

Supposing that there is an efficient interaction between the discs winds and the BLR clouds, 
we would speculate above effects of interactions. First, the BLR clouds cannot be destroyed 
by the disc winds in the ablation timescale, but the drag force will be significantly enhanced (i.e., 
strongly affect $\zeta_0$). For simplicity, we estimate the wind density
$n_{\rm w}=\dot{M}_{\rm w}/4\pi R_{\rm BLR}^2V_{\rm w}m_{\rm p}$, where $\dot{M}_{\rm w}$ is the wind 
mass rate and $V_{\rm w}$ is its velocity. Using the $R_{\rm BLR}-L$ relation\cite{Bentz2013}, we 
have $n_{\rm w}\approx 10^6\,\dot{M}_{0}M_8^{-1/2}L_{44}^{-3/4}{\rm\,cm^{-3}}$ and thus 
$t_{\rm abl}\approx 2.5\times 10^4\,R_{14}n_{10}\dot{M}_0^{-1}L_{44}$\,yrs, where 
$\dot{M}_{0}=\dot{M}_{\rm w}/1\sunm\,{\rm yr}^{-1}$ and 
$V_{\rm w}$ is taken to be the Keperian velocity at $R_{\rm BLR}$. This is much longer than the 
timescales of capturing clumps. On the other hand, the thermal instability creates new clouds in 
the thermal times scales\cite{Beltrametti1981}. The BLR is covered by more clouds so that the
observed EW of broad-lines will be enhanced significantly. Observational tests could be
done by systematically comparing EW(H$\beta$) of the AGNs with/without BLR outflows. Second, it 
has been suggested that 
superluminous transients might appear in AGNs if disc winds collide with the BLR clouds (the BLR 
clouds are treated as continuous fluid)\cite{Moriya2017}. We should keep cautious about these 
predictions and hope future numerical simulations help understand such a complicated BLR.

\subsection{Implications of the present model}
Given the ionisation parameter of $\Xi=L_{\rm ion}/4\pi \rblr^2c p_{\rm gas}$, we have 
$\rblr\propto L_{\rm ion}^{1/2}\propto L_{5100}^{1/2}$ if photoionization dominates the 
physics of the BLR, where $L_{\rm ion}$ and $L_{5100}$ are the ionizing (for hydrogen atoms) 
and 5100\AA\, luminosities, respectively, and
$p_{\rm gas}$ is the gas pressure of ionized clouds in the BLR. The present model explains 
the $R-L$ relation in term of the constant ionization parameter for photoionization\cite{Bentz2013}.
Why the ionisation parameter is essentially constant remains as a long-term open 
question\cite{Baldwin1995,Baskin2014}, but it is worth of investigating if it is caused by the 
evolution of the clumps from their capture locations until they merge with accretion discs 
(discussed in ``Dynamics of tidally disrupted clumps"). In this paper, we emphasize that the 
BLR originates as a natural consequence of the tidal capture of clumps from the torus, which 
is consequently the source of fuel for the accretion disc. 
We point out that AGNs with
extremely high accretion rates have H$\beta$ lags much shorter than objects with the same
luminosity from the $\rblr\propto L^{1/2}$ relation\cite{Du2014,Du2015,Du2016}. Even more 
complicated is the presence of multiple lags in NGC 5548 when it has quite high luminosity 
($\sim 10^{43.5}\ergs$)\cite{Pei2017} (on the other hand, shortened H$\beta$
lags are common among AGNs with high accretion rates, but NGC 5548 only has once shown such 
behaviour in the last 17 monitoring campaigns, indicating that the shortening mechanisms are 
likely different). This at least indicates that there are sub-structures in the 
BLRs\cite{Li2016}. Furthermore, true type 2 AGNs lacking broad emission lines could be explained 
by either their central black holes having low accretion rates\cite{Nicastro2000}, or the black 
holes having very high accretion rates\cite{Ho2012,Miniutti2013}. The former can be easily 
explained by the lack of captured clumps. The later subject involves the BLR structure, which 
the self-shadowing effects governed by the inner discs strongly
influences in AGNs with high accretion rates\cite{Du2016}.

In our model, tidal captures of clumps from the dusty torus determine the planes of the BLR 
clouds, and therefore the poloidal geometry of the BLR, namely, 
$\Theta_{\rm BLR}\approx \Theta_{\rm torus}$ holds for H$\beta$ regions (low-ionisation line 
regions), and follows torus vertical structure. If the BLR thickness can be reliably estimated 
by modeling reverberation mapping data\cite{Pancoast2011,Li2013}, an interesting correlation 
is expected between $\left(H/R\right)_{\rm BLR},\left(H/R\right)_{\rm torus}$ of AGNs and
quasars, providing one observational test of the present scenario of the BLR origin, where 
$H/R$ with subscripts are the relative thickness at the characterized radius of the BLR 
and the torus, respectively.

It is worth doing numerical simulations of the present complex processes for in-depth 
comparisons with observations by making a more detailed physical and self-consistent 
model, which includes the dynamics of interactions with surroundings 
(diffusive medium and even outflows developed from accretion discs), thermodynamics,
and radiation. According to Equation (\ref{mdot}), a different geometry of the torus 
(covering factor $\mathscr{C}$ and angle $\Theta_{\rm torus}$) could lead to a different
environment within $D_{\rm sub}$ and strongly affect the subsequent evolution of captured
clumps. The present model coupling a cloud's dynamics and thermodynamics, 
in principle, would naturally explain the existence of high/low-ionisation regions during the
spiral-in to galactic centres and lead to the observed correlations among properties of different 
lines found in PG quasars\cite{Boroson1992}. The type C clouds could be accelerated by the radiation 
pressure of the accretion discs and provide a potential sources of outflows, which could be related 
with several issues in AGNs. Accretion discs are currently presumed to power AGNs, but how to 
form such structures from a
torus should be investigated along with mergers of BLR clouds. Observationally, future campaigns 
of monitoring AGNs with different asymmetries\cite{Hu2008a,Zamfir2010,Runnoe2016}, optical 
\feii\cite{Hu2008b}, near infrared emission\cite{Koshida2014} and orientations deduced from the 
ratio of core-to-extend radio emission\cite{Wills1986} will help identify the global structure
and origin of the BLR.

\begin{addendum}
\item[Data Availability Statement] The data that support the plots within this paper and other 
findings of this study are available from the corresponding author upon reasonable request.
\end{addendum}

\end{methods}

\vglue 2.5cm
\begin{spacing}{1.7}
\begin{table}
\centering
{Supplementary Table 1\\ A list of parameters involved in the present model}
{\footnotesize
\label{parameters}
\begin{tabular}{llcc}\hline
Parameter & ~~~~~~~~~~~~~~~~~~~~~~~~Physical meanings & Valid ranges & Uncertainties\\ 
          &                               &  & $\left(\left\langle\Delta X/X\right\rangle\right)$ \\ \hline
$\zeta_0$ &$R_{\rm T}^0\omega_{\rm A}^0/v_{\rm A}^0$: determines trajectories of clumps & $0\sim50$ &1.6\\ 
$\gammaA$ & cloud A distribution index: $N_{\rm A}=N_{\rm A}^0v_{\rm A}^{\gammaA}$ & $0-4$ & 0.5  \\
$\xi_{\rm A}$ & $v_{\rm A}^0=\xi_{\rm A}\left(\zeta_0^2+1\right)^{-1/2}v_{\rm K}^0$, $v_{\rm K}^0$ is the Keplerian velocity at $R_{\rm T}^0$.& $<1$ &0.2 \\
$R_0/\Rg$ & the circulaized radius of type A clouds & $\sim 6.2\times 10^3$& 0.3 \\
$R_{\rm in}$ & radius of type B clouds merging with accretion disks         & $<R_{\rm BLR}$ & 0.6\\
$\gammaB$ & cloud B distribution index: $N_{\rm B}=N_{\rm B}^0v_{\rm B}^{\gammaB}$ & $0-4$ & 0.5  \\
$f_{\rm B}$& fraction of type B clouds  & $0\sim 1$ & 0.3\\ 
$\xi_{\rm C}$ & $v_{\rm C}^0=\xi_{\rm C}v_{\rm K}^0$, $v_{\rm C}^0$ is the ejection velocity of type C clouds at $R_{\rm T}^0$.& $>1$ &0.3 \\
$f_{\rm C}$& fraction of type C clouds, properties shown by Equation (7) & $0\sim 1$ & 0.6 \\
$i$       & inclination angle of observers, $i = 0^{\circ}$ (edge-on) and $i = 90^{\circ}$(face-on) & $\gtrsim 40^{\circ}$ &0.2 \\ \hline
\multicolumn{4}{c}{Auxiliary Parameters (the fixed values are listed by the numbers in bold fonts)}\\ \hline
$\Theta_{\rm torus}$& half angle of torus given by the $\Theta_{\rm torus}-\Loiii$ relation & Ref.$^{1}$ \\
$\alphaA$ & cloud A velocity index: $v_{\rm A}=v_{\rm A}^0\left(R/R_{\rm T}^0\right)^{-\alphaA}$& ${\bf 0.45}; \,\le0.5$\\
$\alphaC$ & cloud C velocity index: $v_{\rm C}=v_{\rm C}^0\left(R/R_{\rm T}^0\right)^{-\alphaC}$& ${\bf 0.30}; \,\le0.5$\\
$\betaA$  & cloud A angular velocity index: $\omega_{\rm A}=\omega_{\rm A}^0 \left(R/R_{\rm T}^0\right)^{-\betaA}$ & ${\bf 1.8};\,\ge 1.5$ \\
$\gammaC$ & cloud C distribution index: $N_{\rm C}=N_{\rm C}^0v_{\rm C}^{\gammaC}$ & ${\bf 0}; \,\sim 0.5$   \\
$\gamma$  & vertical distribution of clumps in torus: $\propto \left(\cos\thetaC\right)^{\gamma}$& ${\bf 1}; \,\sim 1$\\
$\Gamma$  &$1+\alphaA-\betaA$ &  \\ 
$R_{\rm T}^0/\Rg$& tidal disruption radius of clumps.   &  ${\bf 10^4};\,(R_0,R_{\rm torus})$     \\
$R_{\rm BLR}$& the emissivity-averaged radius of the BLR determined by RM.   &   \\
$\varphi_0$  & position angle of tidal event. & $\varphi_0=\pi/2$  \\
$\omega_{\rm A}^0$ & angular velocity of type A clouds at $R_{\rm T}^0$. & absorbed by $\zeta_0$  \\ \hline
\multicolumn{4}{l}{
Note: the averaged $\left\langle\Delta X/X\right\rangle$ is obtained by the statistic of individual $\Delta X/X$ of the sample. Here $X$ is any one}\\ 
\multicolumn{4}{l}{of the fitting parameters in the model.}
\end{tabular}}
\end{table}
\end{spacing}

\newpage

\begin{table}
\centering
{Supplementary Table 2\\ Resultant parameters of best-fittings of four quasars}
{\footnotesize
\label{four-fittings}
\begin{tabular}{lcccc}
\hline
Parameter & PG1354+213 & PG2251+113 & PG1351+640 & PG1700+518 \\ 
\hline
$\zeta_0$ & $6.1_{-6}^{+10}$ & $0.71_{-0.5}^{+1.5}$ & $0.02_{-0.02}^{+5}$ & $38_{-31}^{+13}$ \\
$\xi_{\rm A}$ & $0.36_{-0.1}^{+0.3}$ & $0.59_{-0.1}^{+0.2}$ & $0.42_{-0.1}^{+0.1}$ & $0.63_{-0.1}^{+0.1}$ \\
$\gamma_{\rm A}$ & $0.79_{-0.79}^{+1.4}$ & $3.6_{-1.8}^{+0.9}$ & $0.51_{-0.2}^{+0.9}$ & $1.5_{-0.5}^{+0.9}$ \\
$R_0(10^3 R_{\rm g})$ & $3.2_{-0.9}^{+1.2}$ & $2.6_{-0.7}^{+1.0}$ & $1.4_{-0.8}^{+1.5}$ & $7.5_{-2.5}^{+1.5}$ \\
$R_{\rm in}(10^3 R_{\rm g})$ & $0.31_{-0.1}^{+0.2}$ & $0.38_{-0.1}^{+0.1}$ & $0.39_{-0.1}^{+0.1}$ & $3.6_{-0.9}^{+1.3}$ \\
$\gamma_{\rm B}$ & $1.2_{-0.1}^{+0.1}$ & $2.1_{-1.0}^{+0.9}$ & $2.8_{-1.8}^{+0.9}$ & $3.0_{-1.3}^{+0.9}$ \\
$f_{\rm B}$ & $0.95_{-0.6}^{+0.1}$ & $0.78_{-0.2}^{+0.1}$ & $0.73_{-0.1}^{+0.1}$ & $0.24_{-0.1}^{+0.0}$ \\
$\xi_{\rm C}$ & $2.7_{-1.1}^{+0.5}$ & $1.1_{-0.5}^{+0.2}$ & $2.2_{-0.4}^{+0.1}$ & $2.9_{-0.1}^{+0.1}$ \\
$f_{\rm C}$ & $0.02_{-0.02}^{+0.1}$ & $0.18_{-0.1}^{+0.1}$ & $0.04_{-0.01}^{+0.01}$ & $0.44_{-0.1}^{+0.1}$ \\
$i({^\circ})$ & $75_{-13}^{+8}$ & $66_{-9}^{+9}$ & $74_{-9}^{+5}$ & $61_{-1}^{+2}$ \\ \hline 
\multicolumn{5}{c}{Line profiles} \\ \hline
${\cal A}_{\rm H\beta}$ & $4.85$ & $0.15$ & $12.8$ & $2.90$ \\
${\cal S}_{\rm H\beta}$ & $0.30$ & $0.22$ & $1.41$ & $1.07$ \\
${\cal Z}_{\rm H\beta}$ & $1.34$ & $1.12$ & $1.77$ & $2.15$ \\ \hline
\multicolumn{5}{l}{
Parameters of all the PG quasars will be provided on request.}
\end{tabular}}

\end{table}

\newpage
\begin{table}
\centering
{Supplementary Table 3\\ Classifications of resultant fittings}
{\footnotesize
\label{fittings}
\begin{tabular}{lcl}\hline
Model & Number &Objects (total of 87 PG quasars) \\ \hline
B     & 1      & 1354+213\\
B+C   & 1      & 2251+113\\
A+B   & 12     & 0003+199, 0934+013, 1004+130, 1012+008, 1103-006, 1211+143, 1351+236\\
      &        & 1351+640, 1512+370, 1534+580, 1617+175, 2233+134\\
A+B+C & 73     & 0003+158, 0007+106, 0026+129, 0043+039, 0049+171, 0050+124, 0052+251\\ 
      &        & 0157+001, 0804+761, 0838+770, 0844+349, 0921+525, 0923+129, 0923+201\\ 
      &        & 0947+396, 0953+414, 1001+054, 1011-040, 1022+519, 1048-090, 1048+342\\ 
      &        & 1049-006, 1100+772, 1114+445, 1115+407, 1116+215, 1119+120, 1121+422\\ 
      &        & 1126-041, 1149-110, 1151+117, 1202+281, 1216+069, 1226+023, 1229+204\\ 
      &        & 1244+026, 1259+593, 1302-102, 1307+085, 1309+355, 1310-108, 1322+659\\ 
      &        & 1341+258, 1352+183, 1402+261, 1404+226, 1411+442, 1415+451, 1416-129\\
      &        & 1425+267, 1426+015, 1427+480, 1435-067, 1440+356, 1444+407, 1448+273\\
      &        & 1501+106, 1519+226, 1535+547, 1543+489, 1545+210, 1552+085, 1612+261\\
      &        & 1613+658, 1626+554, 1700+518, 1704+608, 2112+059, 2130+099, 2209+184\\ 
      &        & 2214+139, 2304+042, 2308+098\\ \hline
\multicolumn{3}{l}{
B: $f_{\rm A,C}\le 5\%$; 
B+C: $f_{\rm A}\le5\%$;
A+B: $f_{\rm C}\le5\%$;
A+B+C: $f_{\rm A,B,C}>5\%$.}\\
\multicolumn{3}{l}{If it is less than 5\%, we consider that component to be negligible. 
There are only two profiles with B} \\
\multicolumn{3}{l}{and (B+C) components, only 12 objects with (A+B), while
the majority of 73 objects employ (A+B+C).} \\
\multicolumn{3}{l}{Generally, type B clouds are necessary
in all objects, type A appears in 85/87 objects, and the type C}\\
\multicolumn{3}{l}{ in 74/87.}
\end{tabular}}
\end{table}

\clearpage
\newpage
\begin{figure}
\centering
\includegraphics[angle=0,width=0.95\textwidth]{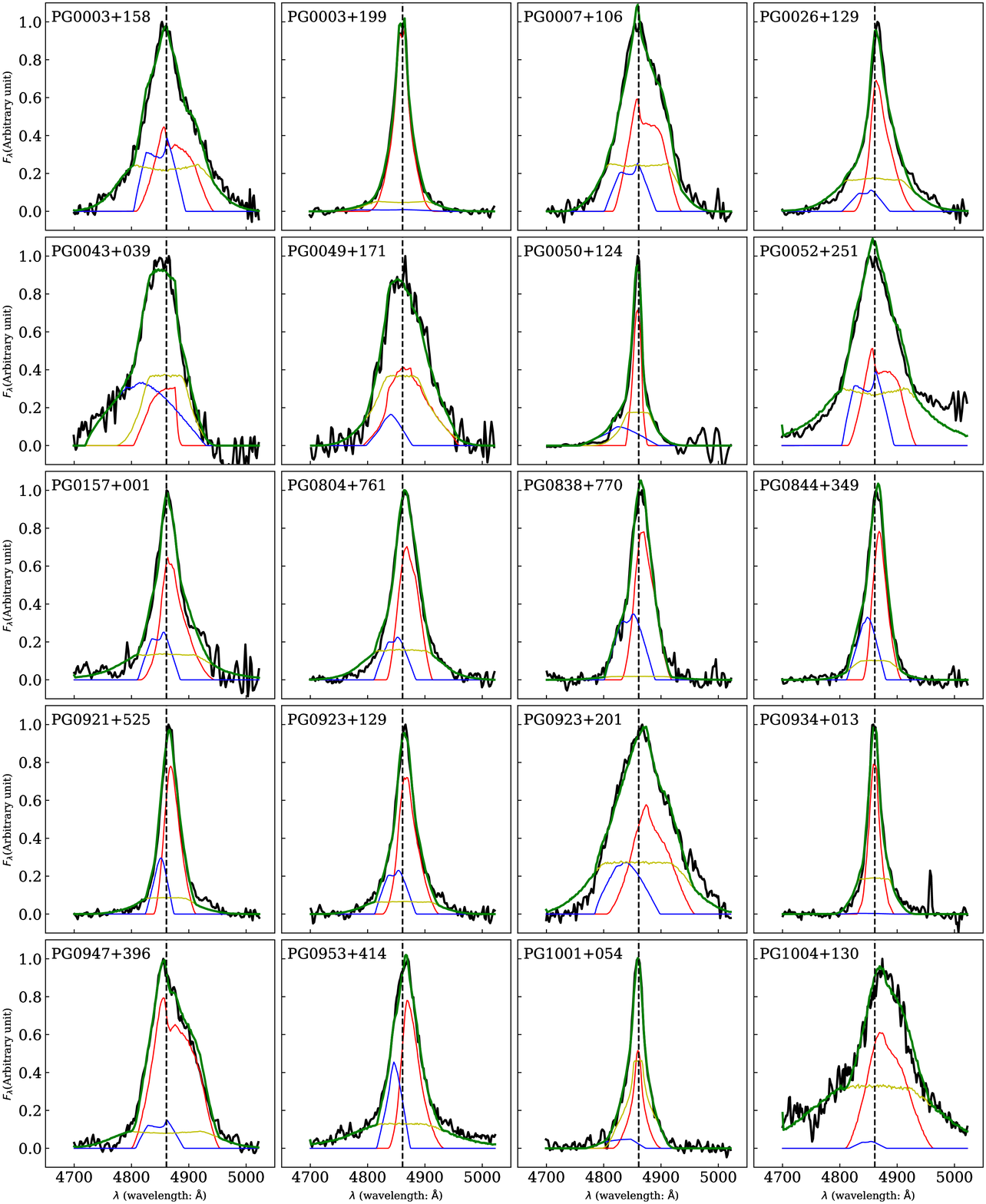}
\vglue -2cm
{\footnotesize Supplementary Figure 1: 
{\bf Comparisons of our model fits with observed H$\beta$ profiles of PG quasars.} Solid 
black lines are the observed profiles, red lines are from type A clouds, the yellow 
from type B and the blue from type C and the green is the total of the three cloud 
types. The dashed lines are $\lambda_0=4861$\AA.
}
\end{figure}

\clearpage
\newpage
\begin{figure}
\centering
\includegraphics[angle=0,width=1\textwidth]{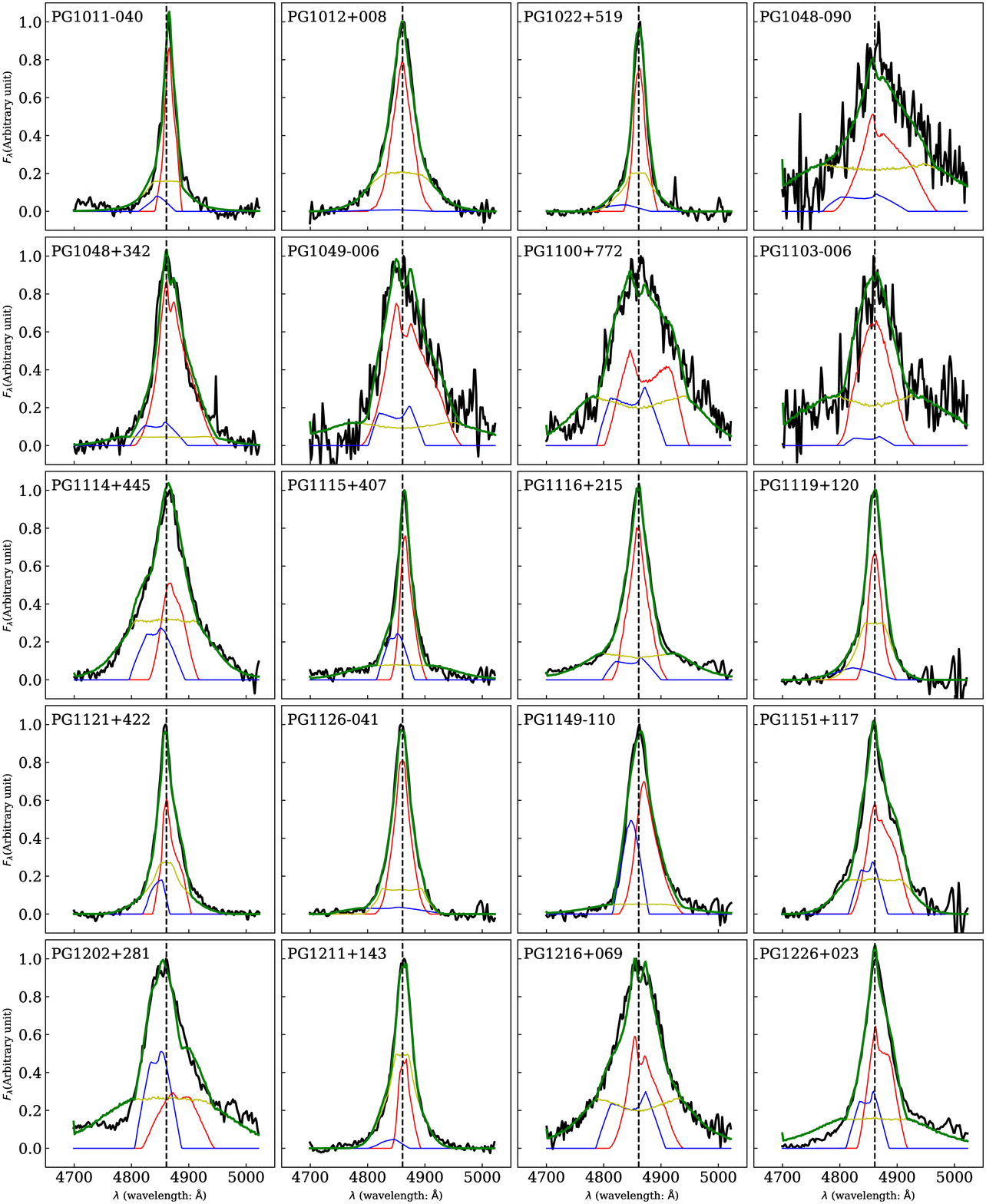}
\vglue -0.8cm
{Supplementary Figure 1 {\it Continued}.}
\end{figure}

\clearpage
\newpage
\begin{figure}
\centering
\includegraphics[angle=0,width=1\textwidth]{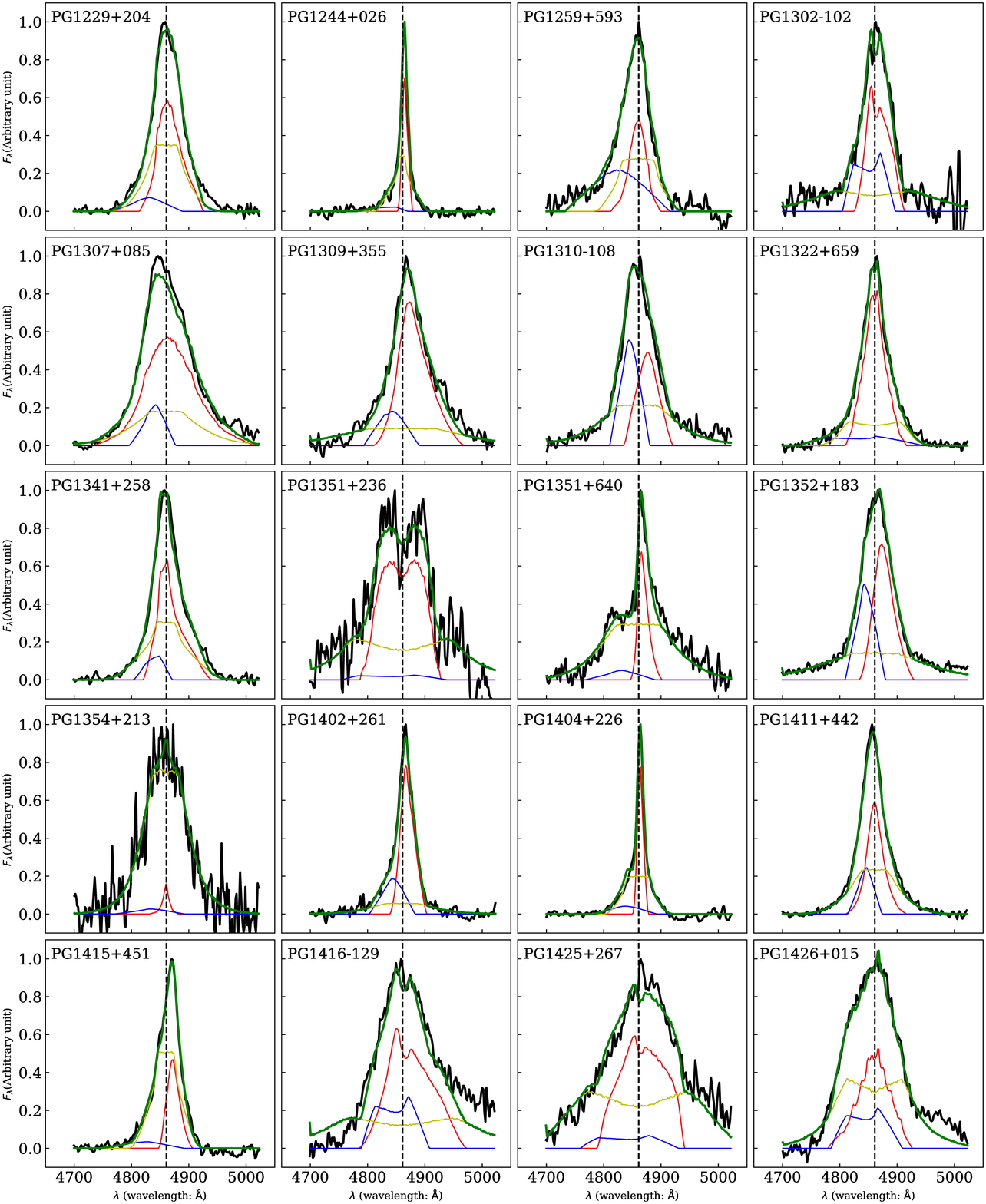}
\vglue -0.8cm
{Supplementary Figure 1 {\it Continued}.}
\end{figure}

\clearpage
\newpage
\begin{figure}
\centering
\includegraphics[angle=0,width=1\textwidth]{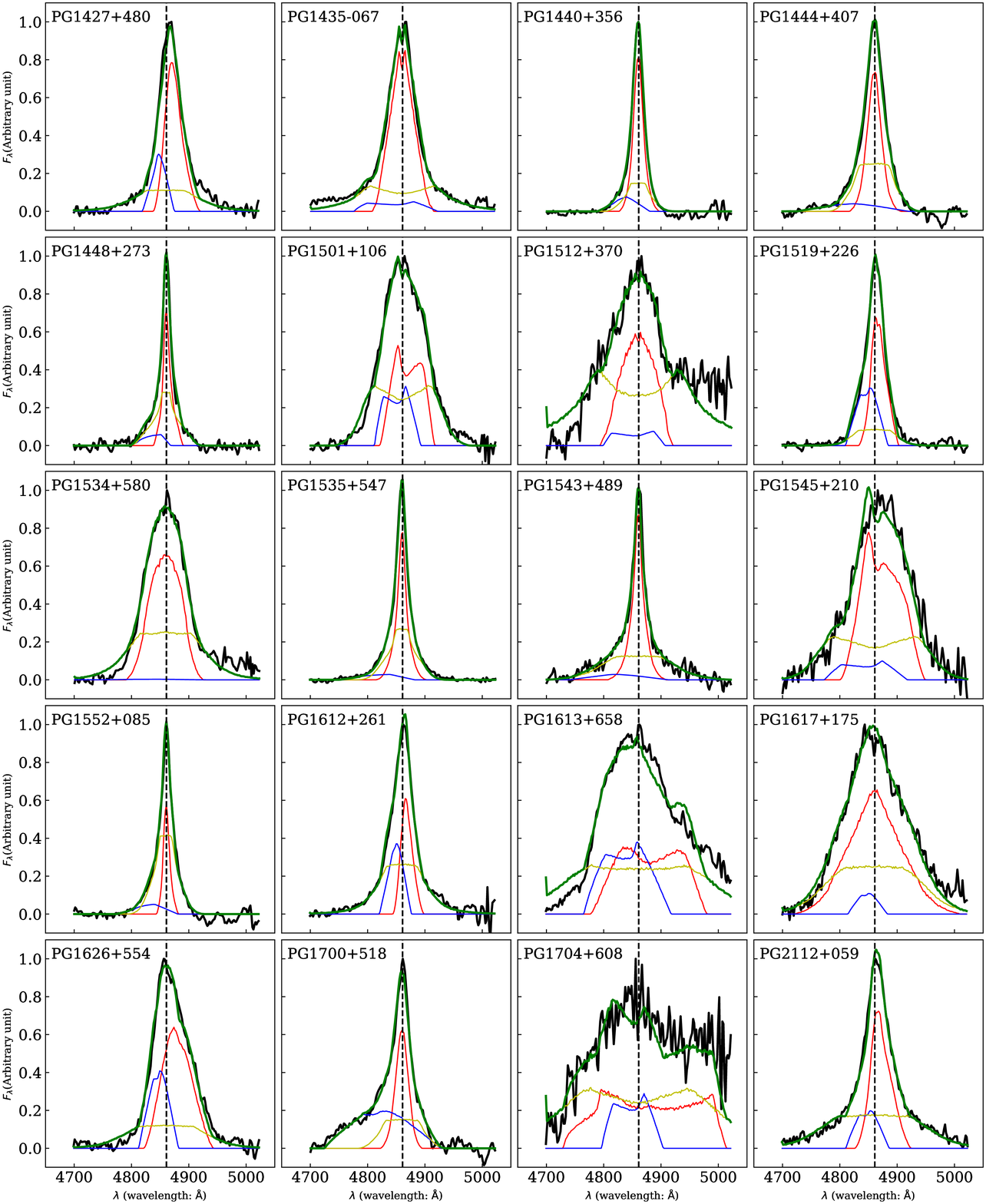}
\vglue -0.8cm
{Supplementary Figure 1 {\it Continued}.}
\end{figure}

\clearpage
\newpage
\begin{figure}
\centering
\includegraphics[angle=0,width=1\textwidth]{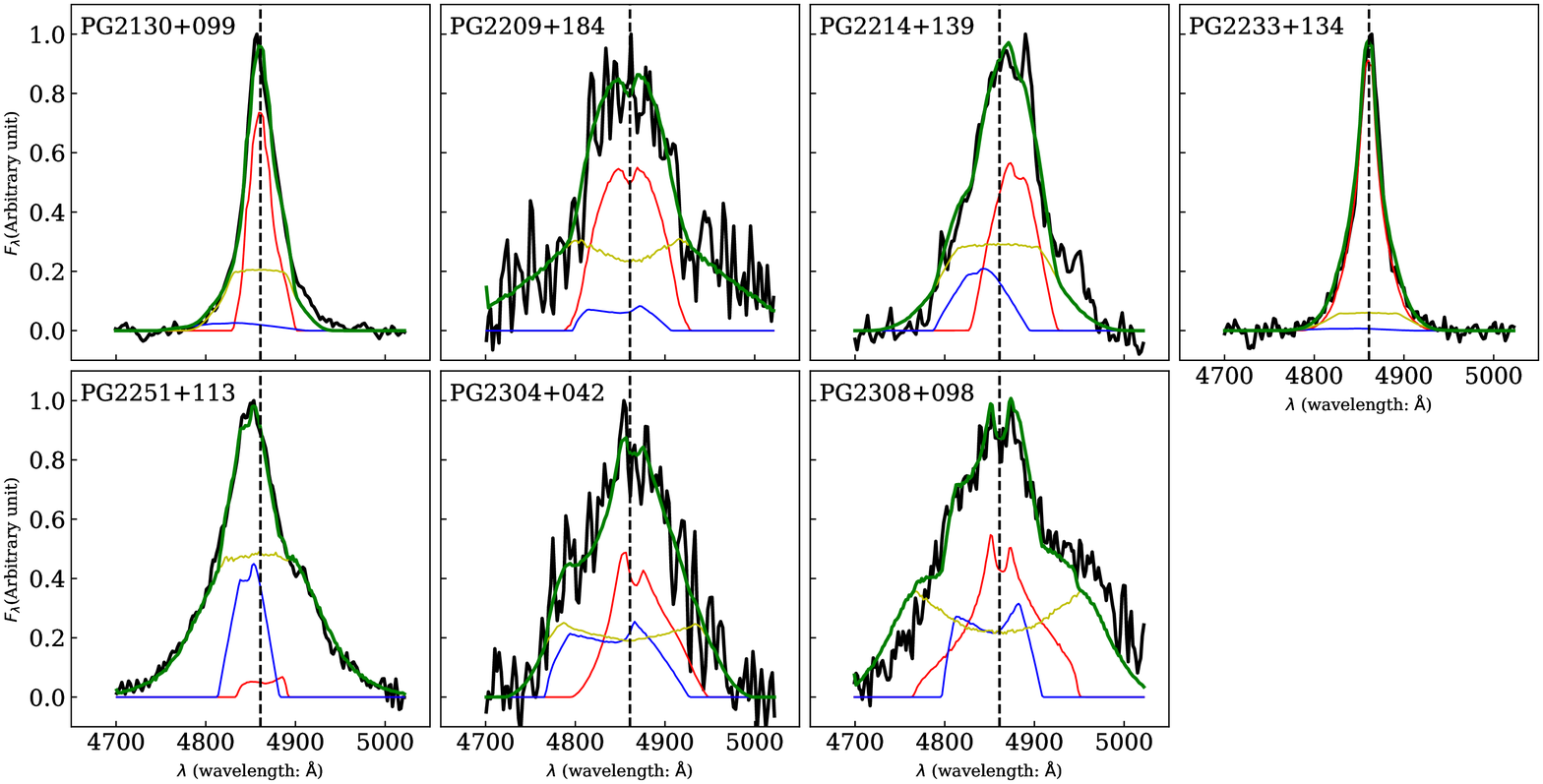}
\vglue -10.8cm
{Supplementary Figure 1 {\it Continued}.}
\end{figure}

\clearpage
\newpage
\begin{figure}
\centering
\includegraphics[angle=0,origin=c,trim=60pt 90pt 0pt 80pt, width=1.05\textwidth]{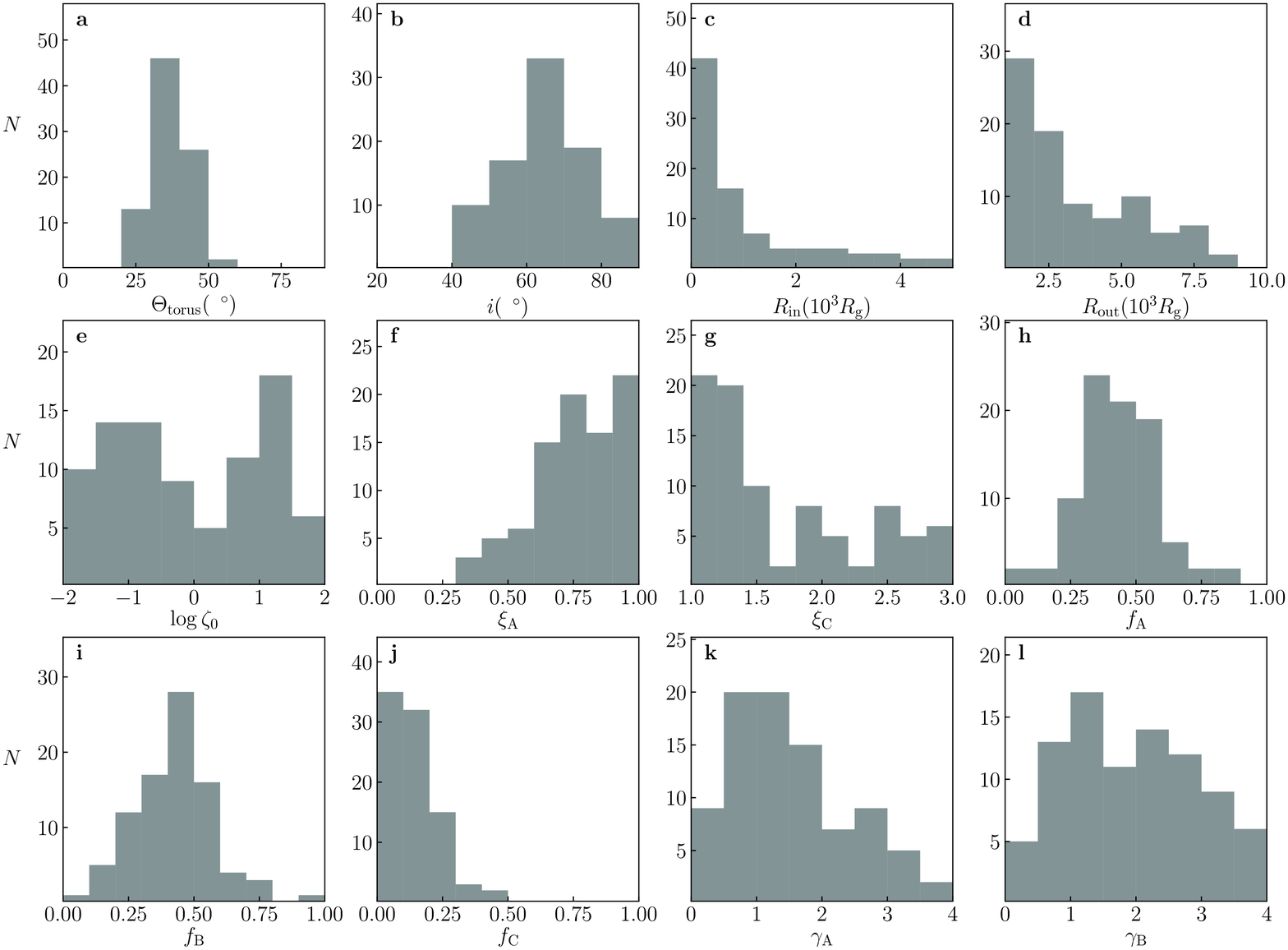}
\vglue 0cm
{\footnotesize Supplementary Figure 2:
{\bf Distributions of the parameters obtained by fitting H$\beta$ profiles.} Note 
$f_{\rm C}=1-(f_{\rm A}+f_{\rm B})$, the average fraction of type C clouds is 
$\langle f_{\rm C}\rangle\approx 0.1$ and $f_{\rm C}\ll (f_{\rm B},f_{\rm A})$ holds for 
most objects of the PG sample. There are a couple of objects with $\xi_{\rm C}\gtrsim 2$, potentially 
implying acceleration of type C clouds driven by radiation pressure, but this possibility needs to
be explored by numerical simulations.
}
\end{figure}

\clearpage
\newpage
\begin{figure}
\centering
\includegraphics[angle=0,origin=c,trim=10pt 90pt 0pt 0pt, width=0.7\textwidth]{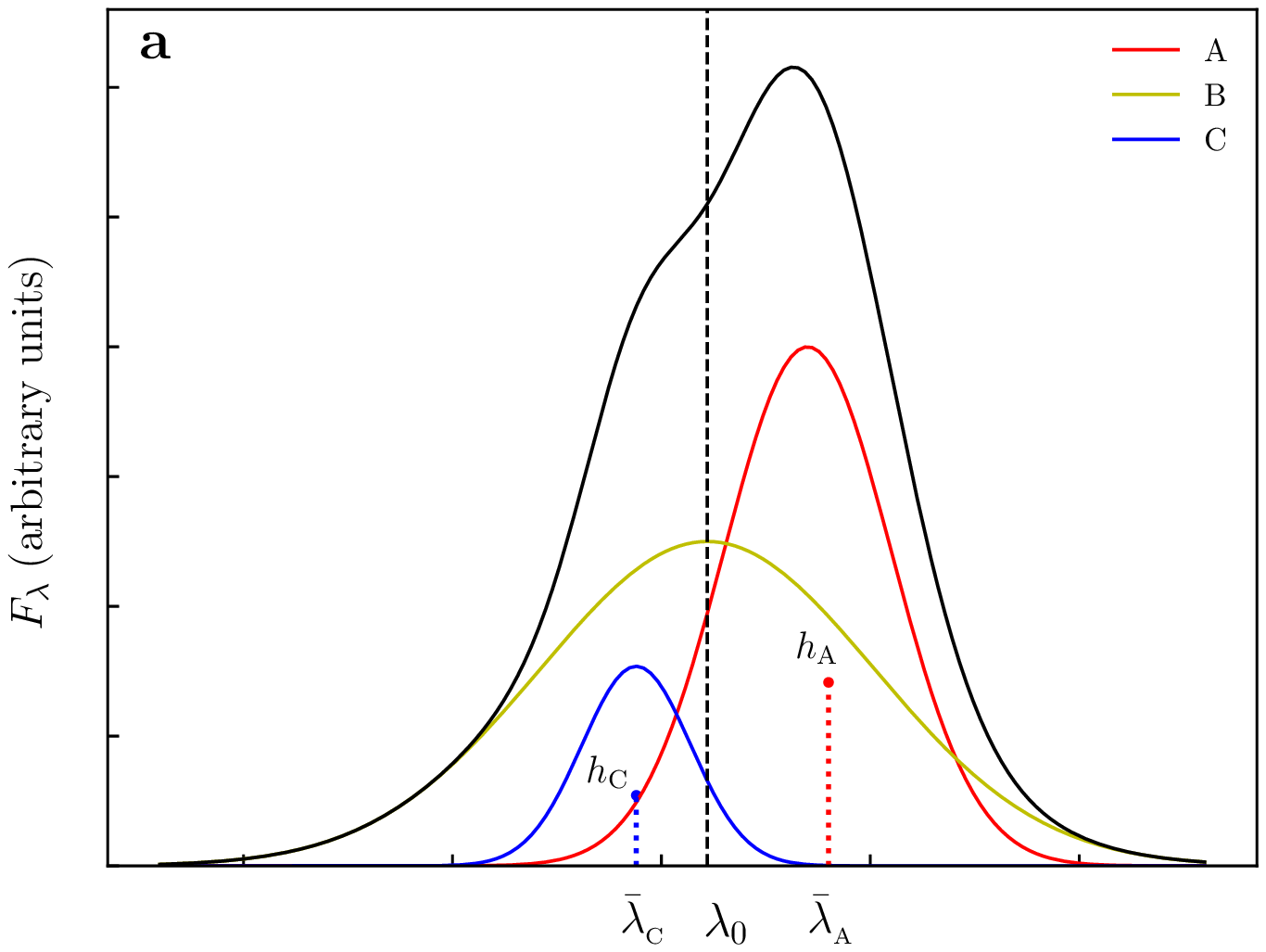}
\vglue -0.7cm
\includegraphics[angle=0,origin=c,trim=10pt 90pt 0pt -100pt, width=0.7\textwidth]{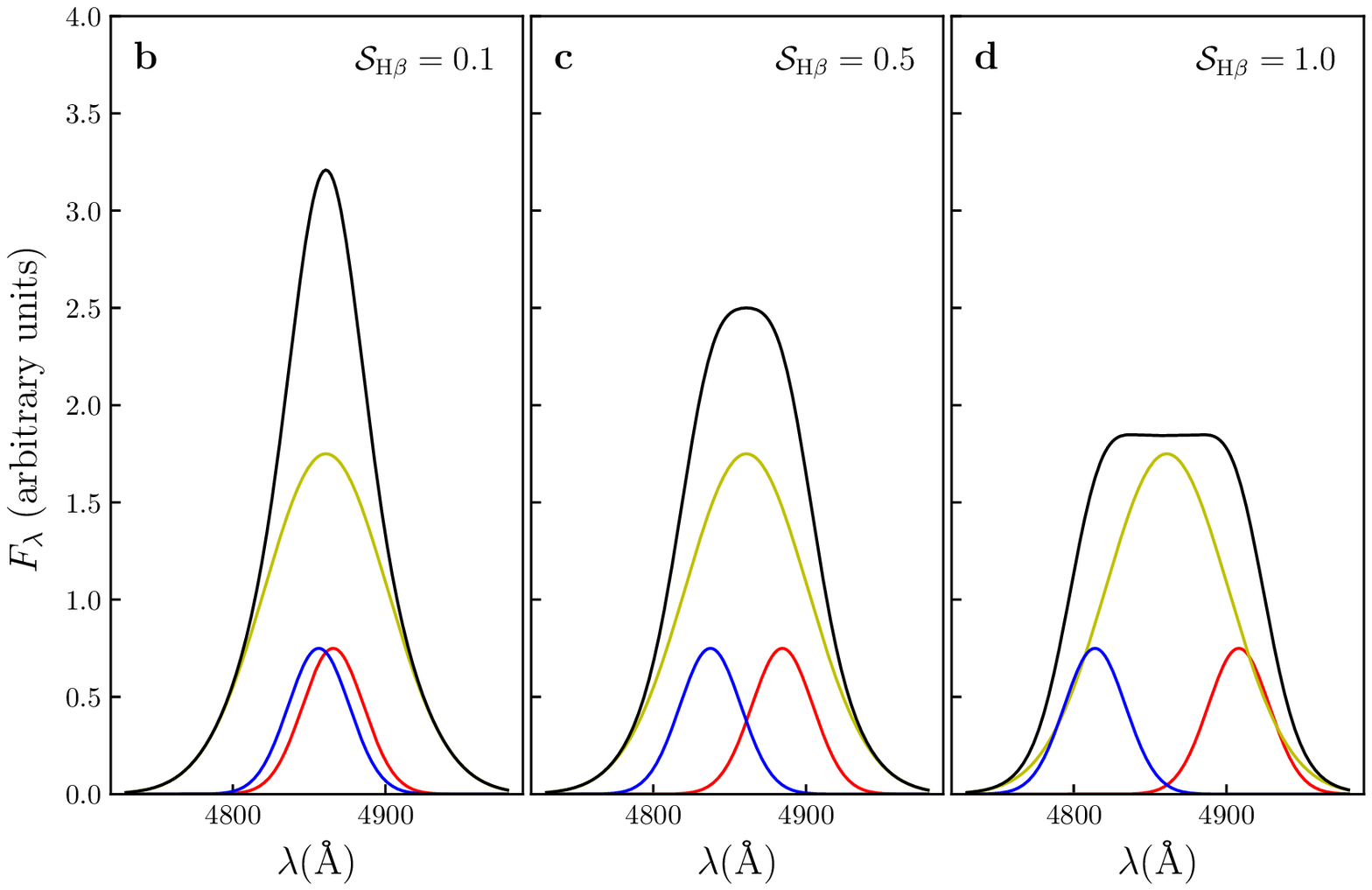}
\vglue 0.6cm
{\footnotesize Supplementary Figure 3: {\bf Illustration of physical meanings of parameters 
describing profiles.} Panel {\it a} shows a cartoon of decomposed profiles for physical meanings 
of the three parameters defined by 
Equation (16). The barycentre wavelength and fluxes are obtained by including 
flux-weight. As the first order approximation, the two parameters of $(h,\bar{\lambda})$ can 
completely describe individual profiles of the decomposed components and hence provide reasonable
asymmetries, shapes and shifts of the total spectra. Panels {\it b-d} illustrate the relationship 
between $\calS$ and real shapes of profiles (from a triangular to boxy).
}
\end{figure}

\clearpage
\newpage
\begin{figure}
\centering
\includegraphics[angle=0,origin=c,trim=10pt 90pt 0pt 0pt, width=0.65\textwidth]{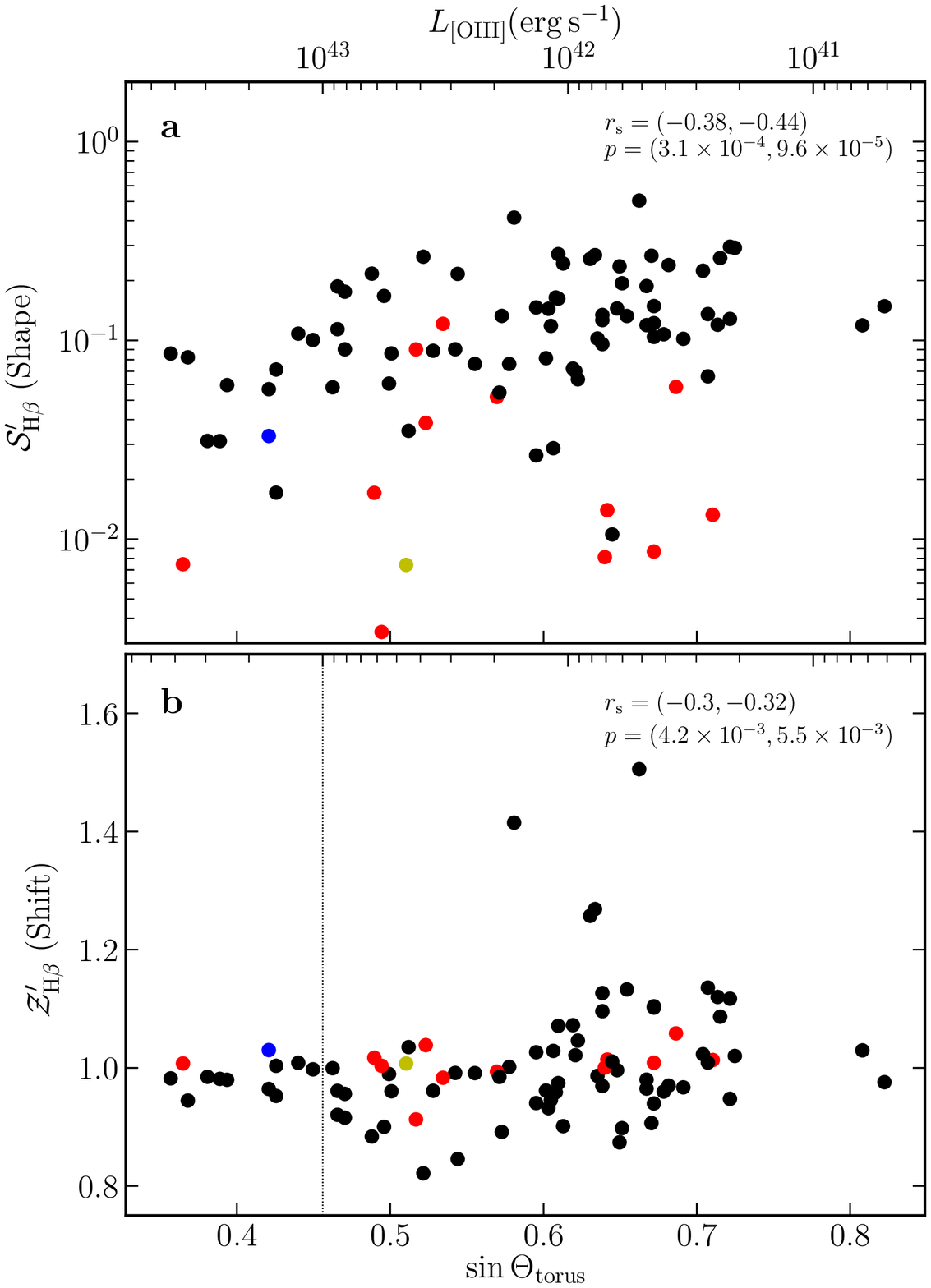}
\vglue 1cm
{\footnotesize Supplementary Figure 4:
{\bf Correlations between $\calS^{\prime}$ and $\calZ^{\prime}$ versus $L_{\rm [OIII]}$}. Panel 
{\rm a} shows a consistent correlation with panel {\it b} in Figure 4. There are 
four outliers with $\calZ^{\prime}\gtrsim 1.2$ deviating from the correlation in panel {\rm b}. 
For objects with $L_{\rm [OIII]}\gtrsim 10^{43}\ergs$, $\calZ^{\prime}\approx 1$ remains, which 
have smaller torus angles ($\Theta_{\rm torus}\lesssim30^{\circ}$).
}
\end{figure}

\clearpage
\newpage
\begin{figure}
\centering
\includegraphics[angle=0,origin=c,trim=10pt 90pt 0pt 0pt, width=0.8\textwidth]{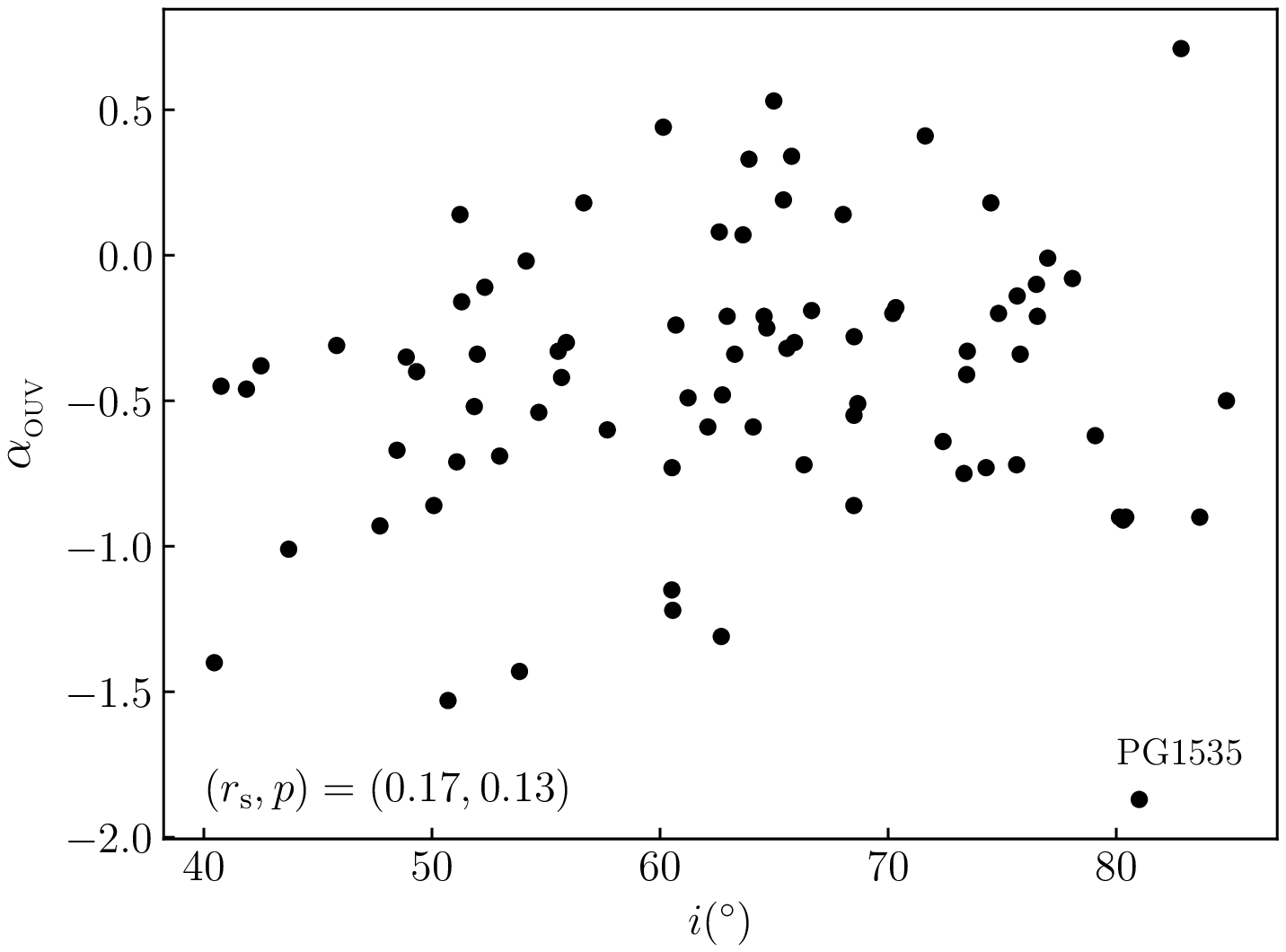}
\vglue 3cm
{\footnotesize Supplementary Figure 5:
{\bf A comparison of PG quasar inclinations from the present model with $\alpha_{_{\rm OUV}}$ as an 
orientation$^{2}$.} This shows that they are consistent with each others.
The Spearman coefficient and null-probability are indicated in the plot except for PG 1535.
Error bars of $\alpha_{\rm OUV}$ are not given by Ref.$^{2}$, and $i$ is given in Supplementary Table 1.
}
\end{figure}

\end{document}